\documentclass[twocolumn,usenatbib]{mnras}

\usepackage{multicol}
\usepackage{graphicx}
\usepackage{bookmark}
\usepackage{threeparttablex}
\usepackage{tabularx}
\usepackage[T1]{fontenc}
\usepackage{ae,aecompl}
\usepackage{physics}
\usepackage{pdflscape}
\usepackage{adjustbox}
\usepackage{savesym}
\savesymbol{iint}
\usepackage{graphicx}	% Including figure files
\usepackage{amsmath}	% Advanced maths commands
\usepackage{amssymb}	% Extra maths symbols
\usepackage{etoolbox}
\makeatletter
\patchcmd\@combinedblfloats{\box\@outputbox}{\unvbox\@outputbox}{}{\errmessage{\noexpand patch failed}}
\makeatother
\title[]{Blue Straggler Populations of Seven Open Clusters with Gaia DR2}

\author[K. Vaidya et al.]{
Kaushar Vaidya$^{1}$,\thanks{E-mail: kaushar@pilani.bits-pilani.ac.in}
Khushboo K. Rao$^{1}$,
Manan Agarwal$^{1}$,
Souradeep Bhattacharya$^{2}$\
\\
$^{1}$Department of Physics, Birla Institute of Technology and Science - Pilani, 333031, Rajasthan, India\\
$^{2}$European Southern Observatory, Karl-Schwarzschild-Str. 2, 85748, Garching, Germany\\
}

\date{Accepted 2020 June 8. Received 2020 June 3; in original form 2020 January 8}

\pubyear{2020}

\begin{document}
\label{firstpage}
%\pagerange{\pageref{firstpage}-\pageref{lastpage}}
\maketitle

\begin{abstract}
Blue straggler stars (BSS) are well studied in globular clusters but their systematic study with secure membership determination is lacking in open clusters. We use Gaia DR2 data to determine accurate stellar membership for four intermediate-age open clusters, Melotte 66, NGC 2158, NGC 2506 and NGC 6819, and three old open clusters, Berkeley 39, NGC 188 and NGC 6791, to subsequently study their BSS populations.  The BSS radial distributions of five clusters, Melotte 66, NGC 188, NGC 2158, NGC 2506, and NGC 6791, show bimodal distributions, placing them with Family II globular clusters which are of intermediate dynamical ages.  The location of minima, $r_\mathrm{{min}}$, in the bimodal BSS radial distributions, varies from 1.5$r_c$ to 4.0$r_c$, where $r_c$ is the core radius of the clusters. 
We find a positive correlation between $r_\mathrm{{min}}$ and $N_{\mathrm{relax}}$, the ratio of cluster age to the current central relaxation time of the cluster.  We further report that this correlation is consistent in its slope, within the errors, to the slope of the globular cluster correlation between the same quantities, but with a slightly higher intercept. This is the first example in open clusters that shows BSS radial distributions as efficient probes of dynamical age. The BSS radial distributions of the remaining two clusters, Berkeley 39 and NGC 6819, are flat. The estimated $N_{\mathrm{relax}}$ values of these two clusters, however, indicate that they are dynamically evolved.  Berkeley 39 especially has its entire BSS population completely segregated to the inner regions of the cluster. 
\end{abstract}


%%%%%%%%%%%%%%%%%%%%%%%%%%%%%%%%%%%%%%%%%%%%%%%%%%

%%%%%%%%%%%%%%%%% BODY OF PAPER %%%%%%%%%%%%%%%%%%

\section{Introduction} \label{Introduction}
Blue Straggler Stars (BSS)  are intriguing objects found in diverse environments such as globular clusters \citep{sandage53, fusi92, sarajedini93}, open clusters \citep{johnson55,burbidge58,sandage62,al95}, OB associations \citep{mathys87}, Galactic fields \citep{preston00}, and dwarf galaxies \citep{momany07,mapelli09}. On color-magnitude diagrams (CMDs) of star clusters, BSS are found along the extension of the main-sequence as the brighter and bluer objects than the main-sequence turn off point of the cluster \citep{sandage53}.  Their presence at these locations on the CMDs suggests that processes that increase the initial masses of stars are at work in these clusters. Observational evidences of BSS being more massive than other member stars of the clusters have been reported \citep{shara97,gilli98,ferraro06,fioren14}. Mass transfer in binary systems \citep{mccrea64}, and stellar mergers resulting from direct stellar collisions \citep{hills76} have been considered as the two chief mechanisms for the formation of the BSS.

BSS are considered crucial probes to study the interplay between stellar evolution and stellar dynamics \citep{bailyn95}. 
Being the most massive objects of the clusters, their radial distributions are expected to show observational signatures of the dynamical friction working at drifting these massive objects to the innermost regions of the cluster.  \citet{ferraro12} discovered that the BSS radial distributions of 21 coeval globular clusters fall into one of the three distinct  families - flat (family I), bimodal (family II), and centrally peaked (family III). As illustrated in \citet{ferraro12}, clusters with centrally peaked BSS radial distributions are dynamically most evolved with all of their most massive members already in the innermost regions. Those with the bimodal BSS radial distributions are of the intermediate dynamical ages where the minima in the distribution, $r_\mathrm{{min}}$, delineates the cluster region up to which the dynamical friction has been effective as yet.  Finally clusters with the flat radial distributions are the least dynamically evolved clusters in which the dynamical friction has not yet affected the initial spatial distribution of the BSS. \citet{ferraro12} further reported a strong correlation between the relaxation times of the family II clusters and the location of $r_\mathrm{{min}}$, which further substantiated their interpretation of BSS radial distributions as indicators of the dynamical ages of the clusters.  

For the first time, \citet{bhattacharya19} reported reliable BSS candidates of open cluster, Berkeley 17 \citep{phelps97}, identified using Gaia DR2 data. They found a bimodal radial distribution of the BSS, similar to family II globular clusters, which are of intermediate dynamical ages. This open cluster was already found to show the effect of mass segregation \citep{bhattacharya17} and hence known to be a dynamically evolved cluster. \citet{bhattacharya19} showed the first example of how the use BSS radial distributions can allow us to make comparison of the dynamical ages of open clusters with those of the globular clusters.  Recently \citet{rain19} presented BSS population of another old open cluster Collinder 261 using Gaia DR2 data and radial velocity data from FLAMES@VLT.  They found a flat BSS radial distribution implying that Collinder 261 is a dynamically young cluster. It is important to study BSS radial distributions in more open clusters to know whether open clusters show the same three families of BSS radial distributions. Globular clusters and open clusters differ particularly in terms of age, metallicity, and stellar density.  Studying BSS populations in open clusters can offer significant insight into the BSS formation scenario and their role in cluster evolution.   

The rest of the paper is organized as follows. Section~\ref{Section 2} gives information about the Gaia DR2 data used in this work,  and the selection of open clusters analyzed here, Section~\ref{Sec. 3} details the method we used for the membership determination, Section~\ref{Sec. 4} gives our results, and finally Section~\ref{Sec. 5} presents a discussion on our results.   

\section{Data and Cluster Selection} \label{Section 2}
We use Gaia DR2 \citep{brown18} data to study the BSS of open clusters.  Gaia DR2 data provides stellar positions RA ($\alpha$) and DEC ($\delta$), proper motions in RA ($\mu_{\alpha}cos\delta$) and in DEC ($\mu_{\delta}$), and parallaxes ($\omega$) with a limiting magnitude of G=21 mag for more than 1.3 billion sources \citep{lindegren18}.  The unique advantage with Gaia DR2 is the unprecedented precision that it offers, e.g., in parallax measurements, up to 0.04 milli-arc-second (mas, hereafter) for G < 15 mag sources, $\sim$0.1 mas for G = 17 mag, and up to 0.7 mas for fainter G > 20 mag sources, in proper motions, up to 0.06 mas yr$^{-1}$ for G < 15 mag, $\sim$0.2 mas yr$^{-1}$ for G = 17 mag, and 1.2 mas yr$^{-1}$ for G > 20 mag. For each open cluster, using the Gaia DR2 proper motions, parallaxes, and radial velocities when available, we identify cluster members including BSS and red giant branch stars (hereafter RGB or reference) populations.    

The most comprehensive catalog of BSS in open clusters containing 1887 BSS candidates of 427 open clusters was developed by \citet[hereafter AL07]{al07}.  They used the photometric data of open clusters through Open Cluster Database $\mathit{WEBDA}$\footnote{https://webda.physics.muni.cz/navigation.html} \citep{mermilliod03}, with membership information in only a small number of clusters. We selected $\sim$90 clusters having $\geq$10 BSS in the AL07 catalog.  In order to obtain more reliable membership information of these BSS populations of clusters, we made use of the Gaia DR2 data to identify cluster members.  Our membership analysis as outlined in Section~\ref{Sec. 3} revealed many inconsistencies in the BSS of various open clusters when compared with the AL07 catalog.  For instance, in several clusters the numbers of BSS as reported in AL07 appeared significantly overestimated when compared with our membership criteria.  \citet{carraro08} had also pointed out similar caveat in the AL07 catalog, namely,
the numbers of BSS tend to correlate with the Galactic latitudes of the given clusters. It is evident that using the Gaia DR2 kinematic information, reliable BSS populations of clusters can be derived.  From our work on the $\sim$90 open clusters, we got around 15 open clusters with reasonably large numbers of BSS ($\geq$15) that would allow a meaningful statistical comparison of their radial distribution with respect to a reference population.  We present here analysis of seven of those clusters.  The analysis of the remaining clusters with $\geq$15 BSS will be presented elsewhere (in preparation Rao et al. 2020).  

Table 1 lists the 7 target clusters whose BSS populations we are presenting in this work.  Four of these clusters, Melotte 66, NGC 2158, NGC 2506, and NGC 6819 are intermediate age $\sim$2-3 Gyr, whereas three clusters, Berkeley 39, NGC 188, and NGC 6791 are old $\sim$6-8 Gyr. The numbers of AL07 BSS candidates in these open clusters range from 15 to 75.  

\begin{table}
%\onecolumn
	\centering
	\caption{Target Open Clusters}
	\begin{tabular}{llll}
		\hline
		\\
	    Cluster& RA & DEC. & No. of BSS \\
		 ~&(deg)&(deg)& (AL07)\\
		 \\
		\hline
		\\
 Berkeley 39& 116.6750 & $-$4.6000 & 43\\
 Melotte 66&111.5958 & $-$47.6666 & 35\\
 NGC 188&11.8666 & $+$85.2550 & 24\\
 NGC 2158 &91.8541 & $+$24.0966 & 40\\
 NGC 2506&120.0041 & $-$10.7700 & 15 \\
 NGC 6791&290.2208 & $+$37.7716 & 75\\
 NGC 6819&295.3250 & $+$40.1866 & 29\\
	\\
		\hline
	\end{tabular}\label{table:A01}
\end{table}

%\begin{center}
\begin{figure*} 
\includegraphics[width=17.5cm]{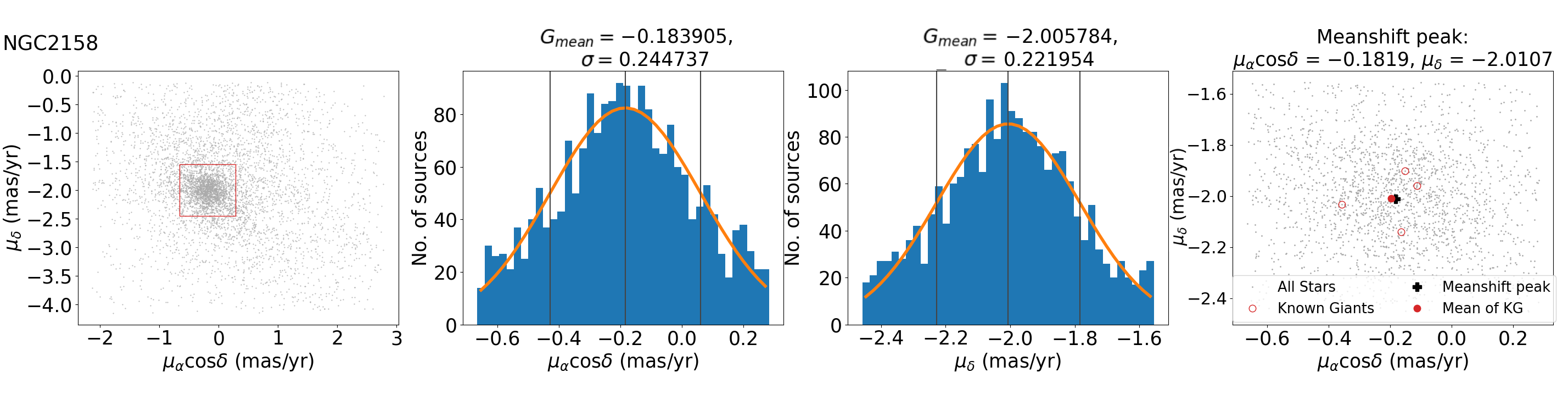}
\caption{The left panel shows a scatter diagram of proper motions of sources within 10$\arcmin$ of the center of NGC 2158.  The rectangular region on this panel shows our initial range of proper motions fixed by visual examination of the distributions.  The two middle panels show the histograms of proper motions in RA and DEC of the selected sources, respectively. The fitted Gaussian functions are overplotted on the distributions. The right panel shows the distribution of proper motions of selected sources, with the proper motions of previously confirmed spectroscopic members shown as red open circles, and their mean position shown with a red filled circle. The peak position calculated from the mean shift algorithm is shown as a black plus symbol. The same figure for the other clusters is given in the Figure~\ref{Fig. A1}.}
\label{Fig. 1}
\end{figure*}
%\end{center}

\begin{table*}
	\centering
	\caption{For each cluster (Column 1), the mean value of the proper motion (RA) determined from the Gaussian 
	fit along with its standard deviation (Columns 2 and 3), the peak value (RA) determined by the mean shift 
	method (Column 4), the mean proper motion (RA) of the previously known confirmed spectroscopic member stars 
	(Column 5), the mean value of the proper motion (DEC) determined from the Gaussian fit along with its 
	standard deviation (Columns 6 and 7), the peak value (DEC) determined by the mean shift method (Column 8), and 
	the mean proper motion (DEC) of the previously known confirmed spectroscopic member stars (Column 9).}
	\adjustbox{max width=\textwidth}{
	\begin{tabular}{rrrrrrrrrrr}
		\hline
		\\
	    Cluster& $\mathrm{G_{mean}}$ (RA)&$\mathrm{\sigma}$ (RA)&$\mathrm{Peak_{ms}}$ (RA)& $\mathrm{Mean_{mem}}$ (RA)& $\mathrm{G_{mean}}$ (DEC)&$\mathrm{\sigma}$ (DEC)&$\mathrm{Peak_{ms}}$ (DEC)& $\mathrm{Mean_{mem}}$ (DEC)  \\
		 ~&(mas/yr)&(mas/yr)&(mas/yr)&(mas/yr)& (mas/yr)&(mas/yr)&(mas/yr)&(mas/yr)\\
		 \\
		\hline
		\\
 Berkeley 39&$-$1.726&0.234&$-$1.728&$-$1.712$\pm${0.081}&$-$1.645&0.173&$-$1.640&$-$1.649$\pm${0.050} \\
 Mellote 66&$-$1.475&0.198&$-$1.479&$-$1.480$\pm${0.074}&~2.740&0.183&2.738&2.732$\pm${0.088}\\
 NGC 188&$-$2.305&0.164&$-$2.300&$-$2.320$\pm${0.148}&$-$0.948&0.146&$-$0.955&$-$0.952$\pm${0.160}\\
 NGC 2158 &$-$0.184&0.245&$-$0.182&$-$0.198$\pm${0.109}&$-$2.006&0.222&$-$2.011&$-$2.009$\pm${0.103}\\
 NGC 2506&$-$2.573&0.191&$-$2.572&$-$2.587$\pm${0.114}&~3.908&0.141&3.904&3.944$\pm${0.126}\\
 NGC 6791&$-$0.422&0.240&$-$0.419&$-$0.438$\pm${0.101}&$-$2.270&0.282&$-$2.274&$-$2.253$\pm${0.113}\\
 NGC 6819&$-$2.907&0.176&$-$2.912&$-$2.919$\pm${0.122}&$-$3.865&0.193&$-$3.867&$-$3.853$\pm${0.122}\\
	\\
		\hline
		\label{table 2}
	\end{tabular}
	}
\end{table*}

\section{Methodology} \label{Sec. 3}
\subsection{Proper Motion and Parallax Ranges for Selection of Member Stars} \label{Sec. 3.1}
In order to determine the proper motion and the parallax ranges for our membership selection criteria, we first downloaded Gaia DR2 data for a small 10$\arcmin$ field around the cluster center for all the clusters. Next, we plotted scatter diagrams of proper motions in RA ($\mu_{\alpha}cos\delta$) versus proper motions in DEC ($\mu_{\delta}$) for each cluster.  By visually examining these plots, we first fixed an initial range of proper motions to select sources whose proper motions appeared distinct from field stars. The left panel in Figure~\ref{Fig. 1} shows this scatter diagram for a representative cluster, NGC 2158. The rectangular region marks our initial ranges of proper motions. Next, we fitted Gaussian distributions to the proper motions of these selected sources and determined the mean ($\mathrm{G_{mean}}$) and the standard deviation ($\sigma$) of the distribution.  The two middle panels of Figure~\ref{Fig. 1} show the Gaussian fitting to the proper motion distributions in RA and DEC, respectively. We also used the mean shift-based clustering algorithm \citep{comaniciu02} to find the peak positions in the distributions of the proper motions of these selected sources.  The right panel in Figure~\ref{Fig. 1} shows the proper motions of the selected sources with this peak value of the proper motions determined using the mean shift method \citep{fukunaga75} marked with a black plus symbol. The proper motions of the confirmed spectroscopic members of the cluster are shown with red open circles and their mean is marked with a red filled circle. The same figure as Figure 1 but for the other clusters is given in Figure~\ref{Fig. A1}. In Table~\ref{table 2}, we list the mean values of the proper motions determined from the Gaussian fitting, $\mathrm{G_{mean}}$, along with their standard deviations, $\mathrm{\sigma}$, the peak values determined by the mean shift method, and the mean proper motion of the previously known confirmed spectroscopic member stars for comparison. For selection of cluster members, we tried seven proper motion ranges, $\mathrm{G_{mean}} \pm 2\mathrm{\sigma}$, $\mathrm{G_{mean}} \pm 2.5\mathrm{\sigma}$, $\mathrm{G_{mean}} \pm 3\mathrm{\sigma}$, $\mathrm{G_{mean}} \pm 3.5\mathrm{\sigma}$, $\mathrm{G_{mean}} \pm 4\mathrm{\sigma}$, $\mathrm{G_{mean}}\pm 4.5\mathrm{\sigma}$, and $\mathrm{G_{mean}} \pm 5\mathrm{\sigma}$. 

For parallax ranges to select member stars, we used the mean Gaia DR2 values of parallaxes of the confirmed spectroscopic  member stars of clusters and their errors.  We tried three parallax ranges using the mean parallax value, $\bar{\omega}$, and the mean in parallax errors, $\Delta \bar{\omega}$, namely, $\bar{\omega}$ $\pm 2 \Delta \bar{\omega}$,  $\bar{\omega}$ $\pm 2.5 \Delta \bar{\omega}$, and  $\bar{\omega}$ $\pm 3 \Delta \bar{\omega}$. 

\begin{figure} 
\centering
\includegraphics[width=\columnwidth]{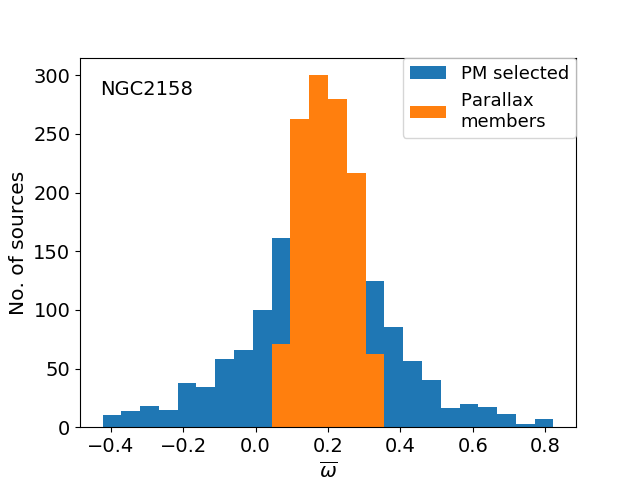} 
\caption{The parallax distribution of proper motion selected sources for the cluster NGC 2158 is shown in blue histogram.  The orange histogram shows the parallax distribution of proper motion selected sources with their parallaxes within the range $\bar{\omega}$ $\pm 3 \Delta \bar{\omega}$, where $\bar{\omega}$ is the mean Gaia DR2 parallax, and $\Delta \bar{\omega}$ is the mean error in the Gaia DR2 parallaxes of the previously known, spectroscopically confirmed members of the cluster.  The same figure showing parallax distributions of the other clusters is given in Figure~\ref{Fig. A2}.}
\label{Fig. 2}
\end{figure}

Along with the Gaia DR2 proper motions and parallaxes, we made use of the previously known confirmed spectroscopic members 
to select our members.  For four of our clusters, radial velocity data under the WIYN Open Cluster Survey (WOCS) have been published earlier which we utilize here. \citet{geller08} presented radial velocity data of more than 1000 sources in NGC 188 confirming membership for 473 sources.  Of these 320 confirmed members are within our adopted radius of the cluster (Section~\ref{Sec. 3.2}).  \citet{tofflemire14} presented the radial velocities for 280 evolved sources of NGC 6791 confirming membership of 111 sources. Most of these confirmed members are in our adopted radius of this cluster (Section~\ref{Sec. 3.2}).  \citet{milliman14} presented the WIYN radial velocity data for 2641 sources in NGC 6819 confirming membership for 679 candidates, almost all within our adopted range of the cluster. \citet{twarog18} presented WIYN radial velocity data for 287 stars in the cluster NGC 2506, confirming membership of 191 sources.  Surprisingly in this cluster, Gaia counterparts of only 26 sources were found.  \citet{bragaglia12} presented VLT/FLAMES spectra for 29 evolved sources in Berkeley 39, all of which are within our adopted radius (Section~\ref{Sec. 3.2}). For the remaining two clusters, Melotte 66 and NGC 2158, spectroscopy for 8 red clump sources were presented by \citet{sestito08} and \citet{smith84,jacobson09}, respectively.

\begin{figure}
\includegraphics[width=\columnwidth]{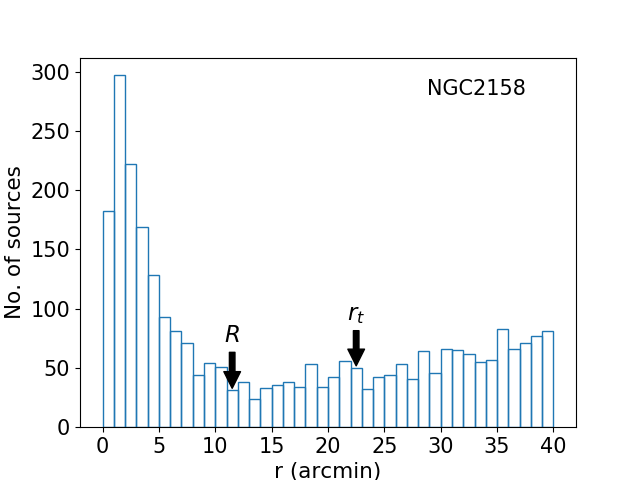}
\caption{The radial distribution of sources with proper motions and parallaxes within the range of selection criteria for the members of the cluster, NGC 2158.  The estimated radius and tidal radius (see Section~\ref{Sec. 4.1}) of the cluster are marked on the figure.  The same figure for other clusters is given in Figure~\ref{Fig. A3}.}
\label{Fig. 3}
\end{figure} 

\begin{table*}
	\caption{The parameters of clusters, age (Column 2), 
	distance (Column 3), metallicity (Column 4), mean extinction in $G$ band (Column 5), mean color excess
	$E(B_P-R_P)$ (Column 6), estimated radius of the cluster (Column 7), core radius (Column 8), 
	tidal radius (Column 9), location of minima in the bimodal BSS radial distribution in units of core 
	radius (Column 10), central relaxation time of the cluster (Column 11), the ratio
	of cluster age to its central relaxation time (Column 12), the literature values of age, distance, and metallicity
	(Columns 13, 14, and 15).} 
%	\centering
	\adjustbox{max width=\textwidth}{
	\begin{tabular}{cccccccccccc|ccc}
		\hline
		\\
		~&~&~&~&~&This Work~&~&~&~&~&~&~&~&Literature&~\\ \hline
	    Cluster & Age & $d$ & Metallicity &$A_G$ &$E(B_P-R_P)$ & Radius & $r_c$ & $r_t$ & $r_{\mathrm{min}}/r_c$ & $t_{rc}$ & $N_{\mathrm{relax}}$ & Age &$d$ & Metallicity \\
	    ~&(Gyr)& (parsec)& ($Z$) & (mag)& (mag) &  ($\arcmin$) & ($\arcmin$) & ($\arcmin$) & ~  & (Myr) & ~ & (Gyr)&(parsec)&($[Fe/H]$)\\
	  
		 \\
		\hline
		\\
 Berkeley 39 & 6.0 & 4254 & 0.0127 & 0.30 & 0.20 & 14 & 1.9 & 20.5 & -- & 78 & 76.9 & 6--8 & 4000 & $-$0.20  \\
 Mellote 66 & 3.4 & 4847 & 0.010 & 0.50 & 0.20 & 15 & 2.9 & 25.1 & 1.68 & 186 & 18.3 & 4--7 & 4700 & $-$0.51 -- $-$0.33 \\
 NGC 188 & 7.0 & 1800 & 0.018 & 0.34 & 0.24 & 30 & 4.1 & 48.7 & 3.25 & 90 & 77.7 & 7 & 1900 & 0.30 \\
 NGC 2158  & 1.9 & 4250 & 0.0186 & 1.0 & 0.53 & 11 & 1.4 & 22.0 & 4.025 & 43 & 44.2 & 1--2 & 4000 & $-$0.63 -- $-$0.3 \\ 
 NGC 2506  & 2.0 & 3110 & 0.008 & 0.23 & 0.10 & 22 & 2.7 & 42.6 & 1.3875 & 150 & 13.3 & 1.85 & 3548 & $-$0.27\\
 NGC 6791 & 8.5 & 4475 & 0.015 & 0.30 & 0.26 & 14 & 2.4 & 24.7 & 2.5 & 216  & 39.3 & 8 & 4000 & 0.31 \\ 
 NGC 6819 & 2.4 & 2652 & 0.019 & 0.20 & 0.20 & 15 & 2.3 & 47.0 & -- & 60 & 40.0 & 2.5 & 2992 & 0.09  \\
 \\
		\hline
		\label{table4}
\end{tabular}
}
\begin{tablenotes}
%\begin{itemize}
\item {Berkeley 39: Age -- \citet{kassis97,kaluzny89}, distance -- \citet{kaluzny89}, metallicity -- \citet{bragaglia12}}
\item {Mellote 66: Age -- \citet{friel93,kassis97}, distance -- \citet{carraro14}, metallicity -- \citet{friel93,sestito08}}
\item {NGC 188: Age --  \citet{sarajedini99}, distance -- \citet{sarajedini99}, metallicity -- \citet{friel10}}
\item {NGC 2158: Age -- \citet{arp62}, distance -- \citet{carraro02}, metallicity -- \citet{carraro02, jacobson09}}
\item {NGC 2506: Age -- \citet{twarog16}, distance -- \citet{twarog16}, metallicity -- \citet{twarog18}}
\item {NGC 6791: Age -- \citet{cunha15}, distance -- \citet{cunha15}, metallicity -- \citet{villanova18}}
\item {NGC 6819: Age -- \citet{rosvick98}, distance -- \citet{kalirai01, balona13,brewer16}, metallicity -- \citet{bragaglia01}}
%\end{itemize}
\end{tablenotes}
\end{table*}

To judge which range of proper motions and parallaxes is most appropriate for selection of cluster members, we considered the retrieval-rate of the previously known confirmed spectroscopic member stars as one criteria, and minimization of the contamination as seen in the CMD as the other criteria. Figure~\ref{Fig. 2} shows proper motion selected sources, and overlaid on them proper motion plus parallax selected sources for a representative cluster, NGC 2158. The same figure as Figure 2 but for the other clusters is given in Figure~\ref{Fig. A2}. In almost all the clusters, for proper motion range, $\mathrm{G_{mean}} \pm 2.5\mathrm{\sigma}$, and for parallax, $\bar{\omega}$ $\pm 3 \Delta \bar{\omega}$, satisfied both of the above criteria. In the last step, we used the cluster radii determined as explained in Section~\ref{Sec. 3.2} to add members to our initial sample of 10$\arcmin$ field to prepare our complete cluster catalogs.

\subsection{Cluster Centers and Cluster Radii} \label{Sec. 3.2}
To estimate the radii of our clusters and use it further to build the complete catalogs of our clusters, we downloaded Gaia DR2 sources for a large field of 40$\arcmin$ radius around the cluster centers.  Next, we plotted radial distributions of the sources to estimate the radius of each cluster.  Figure~\ref{Fig. 3} shows this radial distribution for our representative cluster, NGC 2158. The cluster radius is estimated as the radius at which the cluster radial distribution merges with the field stars distribution.  The same figure as Figure~\ref{Fig. 3} but for the other clusters is given in Figure~\ref{Fig. A3}. The estimated cluster radii are listed in Table~\ref{table4}.  The final catalogs of the clusters are compiled using these values of the cluster radii.   

\begin{figure*}
\includegraphics[width=17.5cm]{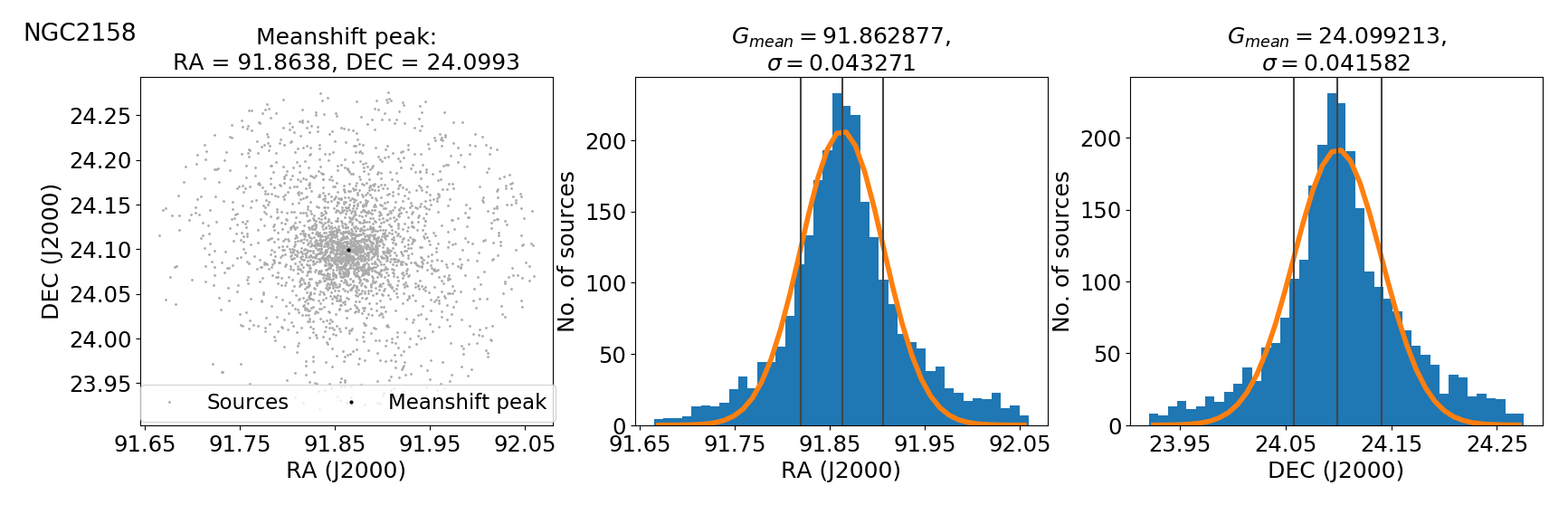}\par
\caption{The left panel shows the result of the mean shift clustering algorithm to determine the cluster center for NGC 2158.  The middle and the right panels show the frequency distributions of cluster members in RA and DEC, respectively, and  Gaussian functions fitted to the distributions.  The same figure with cluster center determinations for the other clusters is given in the Figure~\ref{Fig. A4}.}
\label{Fig. 4}
\end{figure*}

Since the knowledge of accurate cluster center coordinates is important for the analysis that we have carried out, we determined the cluster centers using our final catalogs of clusters following two different methods, the mean shift algorithm \citep{comaniciu02} to determine the densest points in the RA and DEC coordinates, and the fitting of Gaussian functions to frequency distributions of RA and DEC to determine the mean RA and DEC positions.  Figure~\ref{Fig. 4} shows the mean shift algorithm applied to RA and DEC of the cluster NGC 2158 in the left panel, and the Gaussian fits to RA and DEC frequency distributions of the same cluster in the middle and the right panels, respectively.  The same figure as Figure 4 but for the other clusters is given in Figure~\ref{Fig. A4}.  In Table~\ref{table 3}, we give the cluster centers determined by both the methods for all the clusters. 

\begin{table*}
\centering
 \caption{Comparison of cluster center coordinates determined by the mean shift algorithm and fitting of Gaussian functions.}
	\begin{tabular}{ccccc}
		\hline
		\\
		~&Mean Shift&Mean Shift&Gaussian&Gaussian\\ \hline
	    Cluster&RA (deg) &DEC (deg) &RA (deg)&DEC (deg)\\
	  
		 \\
		\hline
		\\
 Berkeley 39 & 116.69925 & $-$04.66973 & 116.70279$\pm$0.05 & $-$04.66830$\pm$0.05 \\
 Mellote 66 & 111.58094 & $-$47.68514 & 111.57993$\pm$0.09 & $-$47.68661$\pm$0.06 \\ 
 NGC 188 & 011.82039 & $+$85.23955 & 011.81136$\pm$1.28 & $+$85.24355$\pm$0.10 \\
 NGC 2158  & 091.86376 &$+$24.09921 & 091.86288$\pm$0.04 & $+$24.09921$\pm$0.04 \\ 
 NGC 2506  & 120.01196 & $-$10.77474 & 120.01064$\pm$0.08 & $-$10.77394$\pm$0.07 \\ 
 NGC 6791 & 290.22079 & $+$37.77634 & 290.21958$\pm$0.07 & $+$37.77634$\pm$0.05 \\ 
 NGC 6819 & 295.33109 & $+$40.18865 & 295.33021$\pm$0.10 & $+$40.18965$\pm$0.07 \\

	\\
		\hline
\label{table 3}
\end{tabular}
\end{table*}

\section{Results} \label{Sec. 4}

\subsection{Radial Density Profiles} \label{Sec. 4.1}
\begin{figure}
\includegraphics[width=\columnwidth]{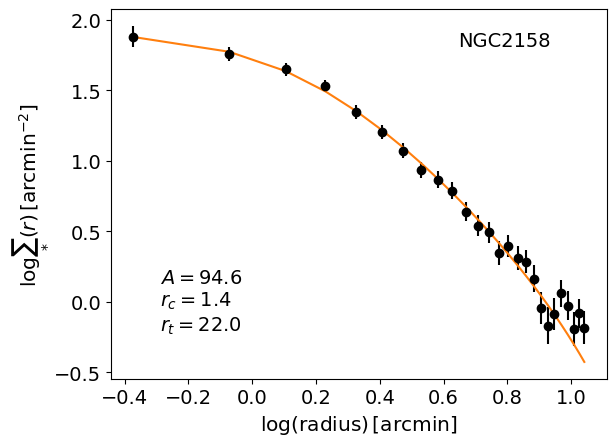}
\caption{The radial density profile of the cluster members is shown with King's function fitting. The error bars represent 1$\sigma$ Poisson errors.  The same figure showing the fitting of the King's function to the radial density profile of the other clusters is given in Figure~\ref{Fig. A5}.}
\label{Fig. 5}
\end{figure}

We plotted radial density profiles of the cluster members using the cluster centers determined as explained in Section~\ref{Sec. 3.2}.  The plot for NGC 2158 is shown in Figure~\ref{Fig. 5}.  We have fitted King's function to the radial distribution of the cluster \citep{kings62}.  In order to fit the King's function, we first divided the range of cluster radius in $\sim$30 equal radius bins.  We then plotted the logarithm of the number density against the radius for each bin.  The King's function is seen to fit well to the clusters (see the Figure~\ref{Fig. A5} in appendix for the remaining clusters).  The resultant parameters, core radii ($r_c$) and tidal radii ($r_t$) are listed in Table~\ref{table4}, and these along with the normalization factor $A$ are as well marked on the figures.

\subsection{Color Magnitude Diagrams} \label{Sec. 4.2}
\begin{figure}
\includegraphics[width=\columnwidth]{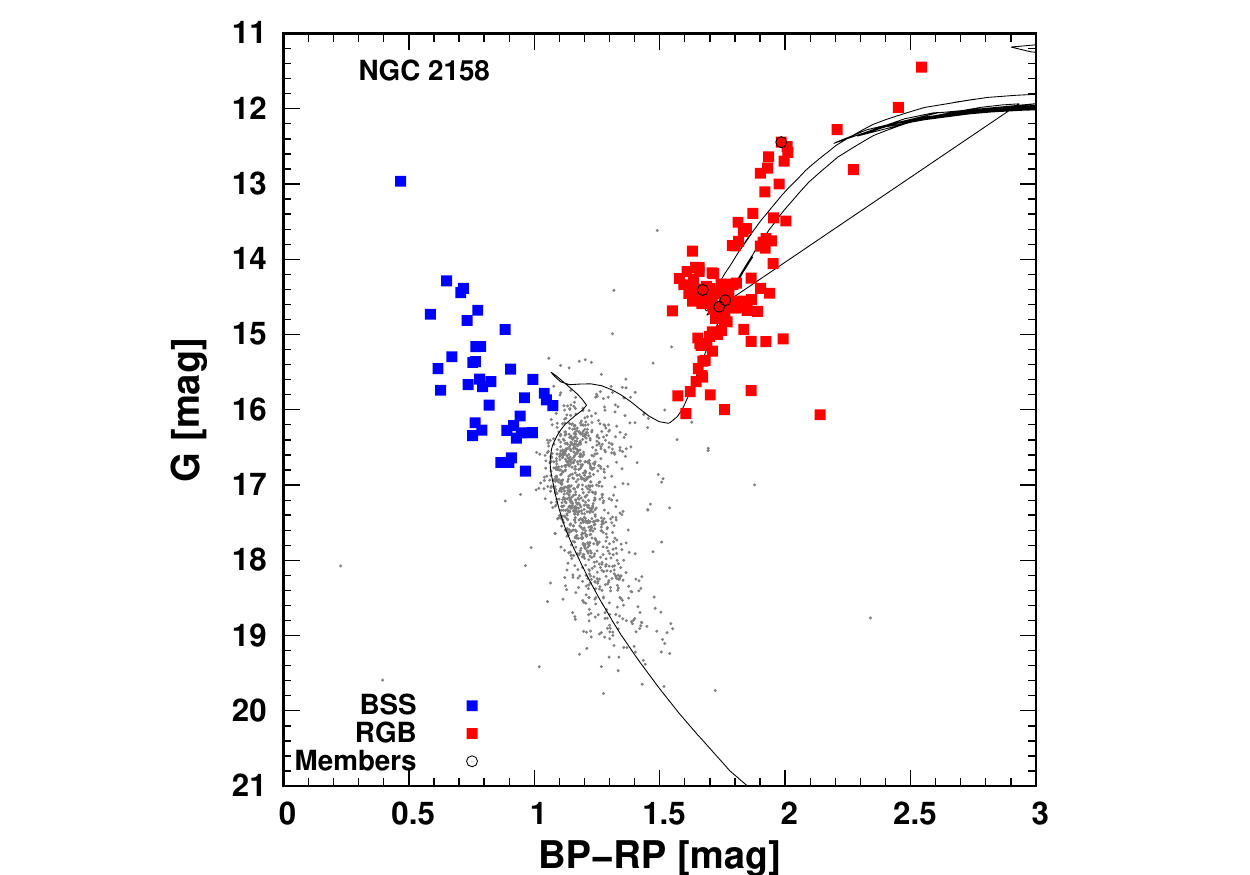}
\caption{The CMD of the cluster NGC 2158 is shown with the fitted PARSEC isochrones. Our BSS are marked as blue solid squares, BSS from AL07 (if any) are marked as green open triangles, and BSS known with spectroscopy data in the literature (if any) are marked as magenta open circles. Previously known spectroscopically confirmed members \citet{jacobson09} and \citet{smith84} are marked as black open circles.  Our RGB populations are marked as red solid squares.  Figure~\ref{Fig. A6} shows the CMDs of the other clusters.}
\label{Fig. 6}
\end{figure}

We plot the CMDs of the clusters using our identified cluster members.  To fit the isochrones, we downloaded PARSEC isochrones \citep{bressan12} of the known ages and metallicity values from the literature (see references in Table~\ref{table4}), and used the mean value of the distances of our bright  ($G$ < 15 mag)  members \citep{bailer18} along with the median values of the extinctions in the $G$ magnitude, $A_G$, and reddening, $E(B_P$-$R_P)$, of our members, available in the Gaia DR2.  For some CMDs, some fine-tunning of parameters was necessary to fit the isochrones to members. In such instances, we varied the metallicity, $A_G$, and $E(B_P$-$R_P)$ values, keeping the distance, and age fixed, to obtain the best fitting of the isochrones. Sources with $G \leq$ TO$_{\mathrm{Mag}}$+0.5, and $B_P-R_P \leq$ TO$_{\mathrm{Col}}-$0.05, where TO$_{\mathrm{Mag}}$ and TO$_{\mathrm{Col}}$ are the magnitude and the color of the main-sequence turn-off point, respectively, were identified as our BSS candidates (blue solid squares). Sources with $G \leq$ TO$_{\mathrm{Mag}}-$0.5, and $B_P-R_P$ redder than the color value of the bottom of the red-giant branch of the PARSEC isochrone were identified as our RGB candidates (red solid squares). The CMD of NGC 2158 is shown in Figure~\ref{Fig. 6}. The previously known spectroscopically confirmed members of the clusters are shown as black open circles.  Figure~\ref{Fig. A6} gives the CMDs of the remaining clusters.  Table~\ref{table4} gives the parameters of the fitted isochrones, and a comparison of the known parameters from the literature.

\subsection{BSS Populations} \label{Sec. 4.3}
NGC 2158 and NGC 6791 are rich in both BSS and RGB populations containing more than 40 BSS candidates each, and 137 and 243 RGB candidates, respectively. NGC 188, NGC 2506, and Berkeley 39 have 20--30 BSS candidates and 55--75 RGB candidates. Melotte 66 contains 14 BSS candidates, the least among our clusters, but contains 106 RGB candidates.  NGC 6819 contains 14 BSS candidates and 68 RGB candidates. In most of the clusters, BSS candidates are up to 2--2.5 mag brighter in the $G$ band from the main-sequence turn-off point.  NGC 2158 and NGC 2506 contain a bright BSS candidate each, which is $\sim$3 mag brighter than the main-sequence turn-off point.  All the clusters except NGC 188 show presence of a significant red clump with confirmed spectroscopic members in several clusters such as Melotte 66, NGC 2158, NGC 2506, and NGC 6791.

\begin{table*}
	\caption{Comparison of the BSS candidates identified in this work with the BSS known in the literature: 
	numbers of BSS candidates and numbers of new BSS candidates identified in this work (Columns 2 and 3), 
	numbers of BSS candidates in the AL07 catalog (Column 4), AL07 BSS candidates that are found members according 
	to our criteria (Column 5), BSS candidates that are common with the AL07 catalog (Column 6), numbers of 
	confirmed BSS in the literature (Column 7), known BSS from the literature that are found members according to 
	our criteria (Column 8), BSS candidates that are common with the confirmed BSS from the literature (Column 9).}
	\begin{tabular}{ccc|ccc|ccc}
		\hline
		\\
		~&This work&~&~&AL07~&~&~&~Literature&\\ \hline
	    Cluster &$N_{BSS}$&New BSS&$N_{BSS}$&Members&Common BSS&$N_{BSS}$ &Members& Common BSS\\
	  
		 \\
		\hline
		\\
 Berkeley 39  & 23 & 9  & 42  & 16 & 14 & -- & -- & -- \\ 
 Mellote 66 & 14 & 5  & 35 & 9 & 9 & -- & -- & --\\
 NGC 188 &  24 & 8  & 24 & 18 & 16 & 20 & 18 & 16 \\
 NGC 2158  & 40 & --  & 40 & -- & -- & -- & -- & -- \\ 
 NGC 2506  & 28 & 22  & 15 & 6 & 6 & -- & -- & --\\
 NGC 6791 & 47 & 32  & 75 & 18 & 15 & 7 & 6 & 6 \\ 
 NGC 6819 & 14 & 2  & 29 & 3 & 2 & 17 & 12 & 12 \\

	\\
		\hline
\label{table 5}
	\end{tabular}
\begin{tablenotes}
\item {NGC 188: \citet{geller08}}
\item {NGC 6791: \citet{tofflemire14}}
\item {NGC 6819: \citet{milliman14}}
\end{tablenotes}
\end{table*}

We compared our BSS candidates with AL07 BSS candidates of the clusters, as well as with the confirmed BSS of the three clusters, NGC 188 \citep{geller08}, NGC 6791 \citep{tofflemire14}, and NGC 6819 \citep{milliman14}.  In AL07, the sources of photometric information are listed in their Table~1 \citep{al07}. We referred to $\mathit{WEBDA}$ and any data linked on ADS\footnote{https://ui.adsabs.harvard.edu} to the references of the photometric data, to obtain the RA and DEC of the AL07 BSS. The comparison of our BSS candidates with AL07 catalogs has only been possible for those sources whose RA and DEC information has been found from the above mentioned resources. For NGC 188, NGC 6791, and NGC 6819, our comparison with the known spectroscopically confirmed BSS populations lends a confirmation to our membership analysis. Using the available spectroscopic information of other member stars of these three clusters, we also ascertain membership of our new, previously unknown BSS candidates in these clusters, and discard them if they are found to be either confirmed or likely non-members.  Table~\ref{table 5} gives a summary of the comparison of our BSS candidates with the previously known BSS.

\subsubsection{NGC\,188} \label{Sec. 4.3.1}
We identified 26 BSS candidates in NGC\,188.  The BSS populations of NGC\,188 have been studied in great detail.  In particular, \citet{geller08} found 20 confirmed BSS based on the radial velocities of sources from WIYN data.  Due to our conservative membership criteria based on Gaia DR2 proper motions and parallaxes, 2 of the \citet{geller08} BSS are not identified as members by us. Of the 18 BSS of \citet{geller08} that we find members, 16 are also our BSS candidates. One BSS of \citet{geller08} is very close to the main-sequence in our CMD, and thus is not included in our BSS candidates. \citet{geller08} also mentioned that this particular BSS (source \#1366) has a low (23\%) membership probability based on its proper motion and is located at 29.6$\arcmin$ from the cluster center. The second BSS from \citet{geller08} appears on RGB in our CMD and has been corrected in the author's later work \citep{geller13}. We found 10 new BSS candidates that are not included in the list of \citet{geller08}. Of these, four sources are listed as binary with unknown membership ``BU'' whereas one source is listed as a single source with unknown membership ``U'' by \citet{geller08}.  Sources with unknown membership in \citet{geller08} lack complete orbit solution in case of binary sources, and lack at least three separate observations of radial velocity with a base-line of one year in case of single sources.  Two new BSS candidates are identified as members by \citet{geller08} but they do not group them into BSS possibly because they are closer to the main-sequence in their CMD. Two of our new BSS candidates are classified as likely non-members or non-members by \citet{geller08}, and for one new BSS candidate there is no information in \citet{geller08}.  We drop the two likely non-members from further analysis but retain remaining eight new BSS candidates which are either members or with candidates with unknown membership or without information in \citet{geller08} in our analysis. This leaves us with 24 BSS candidates in this cluster, 8 of which are new candidates (Table~\ref{table 5}).     

\subsubsection{NGC\,6791} \label{Sec. 4.3.2}
NGC 6791 has 48 BSS candidates, the highest among the clusters that we present in this work.  \citet{al07} had reported 75 BSS in NGC 6791 based on the photometric study of the cluster by \citet{kaluzny90}.  We found Gaia counterparts of all of these sources within 1$'$ search radius of the RA, DEC positions from \citet{kaluzny90}.  However, only 18 of these 75 sources are our members, and 15 are our BSS candidates. \citet{tofflemire14} presented radial velocities of the evolved populations of NGC 6791 in their WIYN open cluster study.  Within our adopted cluster radius, they found 7 BSS which are confirmed members based on their radial velocities.  All but one of these BSS have been identified as BSS candidates in our analysis. From our 42 new BSS candidates which are not identified as BSS by \citet{tofflemire14} (though 9 of them are common 
with AL07), the membership analysis is available for 8 sources in \citet{tofflemire14}.  One of these sources is a confirmed member, 5 are single sources with unknown membership including two rapid rotators, one is a binary with unknown membership, and a single source is a binary likely non-member.  The remaining 34 new BSS candidates lack observations in \citet{tofflemire14} probably as they are below their cut-off magnitude for the targets ($V$=16.8 mag).  After removal of one likely non-member from our BSS list, we are left with 47 BSS candidates in this cluster of which 32 are new candidates as listed in Table~\ref{table 5}.  

\subsubsection{NGC\,6819} \label{Sec. 4.3.3}
We found 18 BSS candidates in NGC\,6819. In the WIYN study of the cluster, \citet{milliman14} found 17 BSS candidates, of which 12 were chosen for detailed Barium abundance study by \citet{milliman15}. According to our membership criteria, 5 of these 17 BSS are not members because their parallaxes are significantly out of the range that we have chosen as membership criteria. The remaining 12 BSS of \citet{milliman14} are our BSS candidates.  We have 6 new BSS candidates, but four of these are likely non-members according to \citet{milliman14}, one source has an unknown membership whereas one has not been observed by \citet{milliman14}.   After discarding the 4 likely non-members, we have a total of 14 BSS candidates in this cluster, of which two are new BSS candidates.
\begin{figure*}
\begin{multicols}{3}
\includegraphics[width=6.0cm]{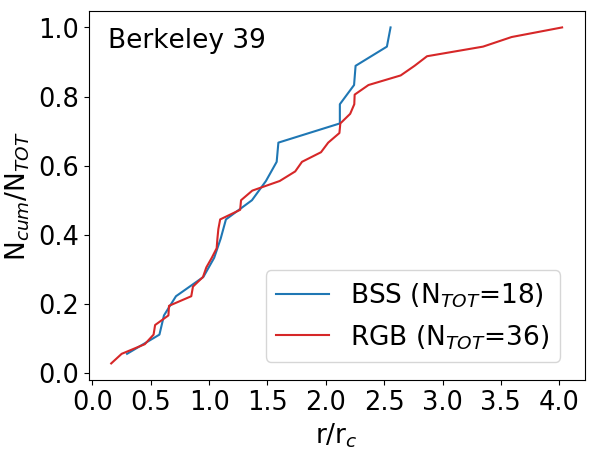}\par
\includegraphics[width=6.0cm]{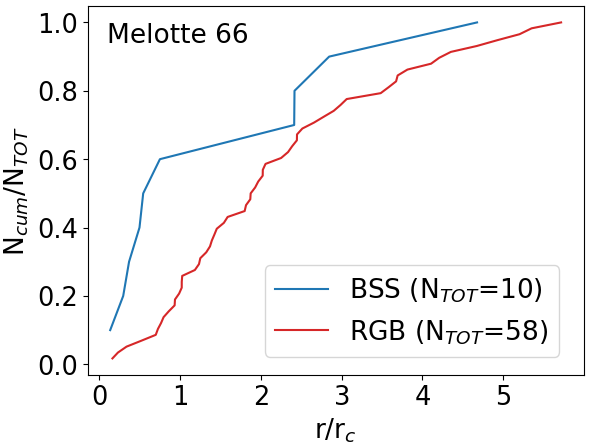}\par
\includegraphics[width=6.0cm]{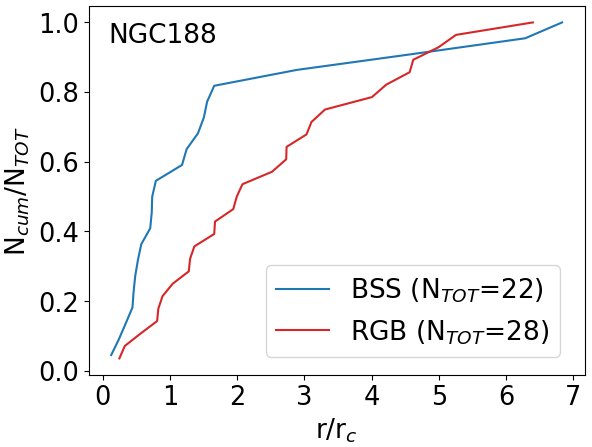}\par
\end{multicols}
\begin{multicols}{3}
\includegraphics[width=6.0cm]{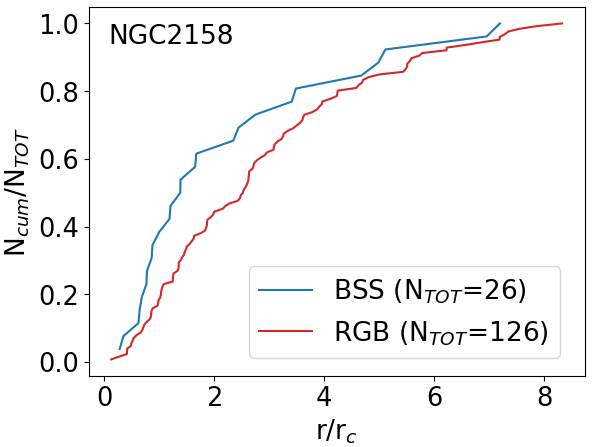}\par
\includegraphics[width=6.0cm]{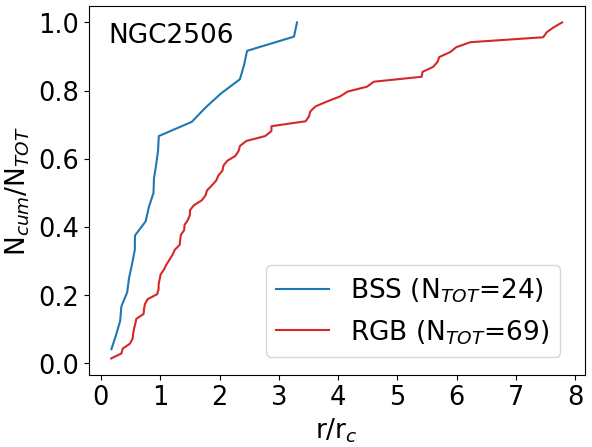}\par
\includegraphics[width=6.0cm]{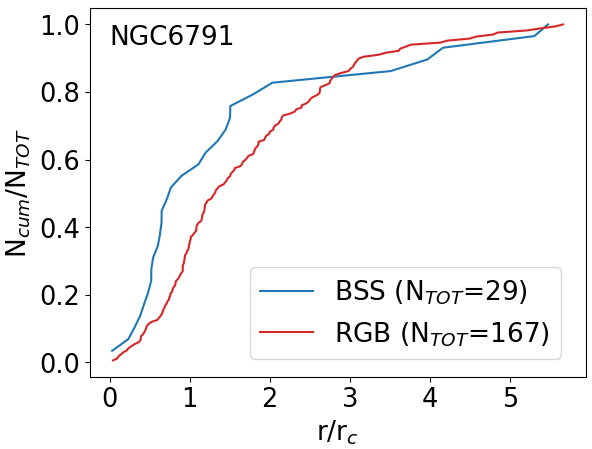}\par
\end{multicols}
\begin{multicols}{3}
\includegraphics[width=6.0cm]{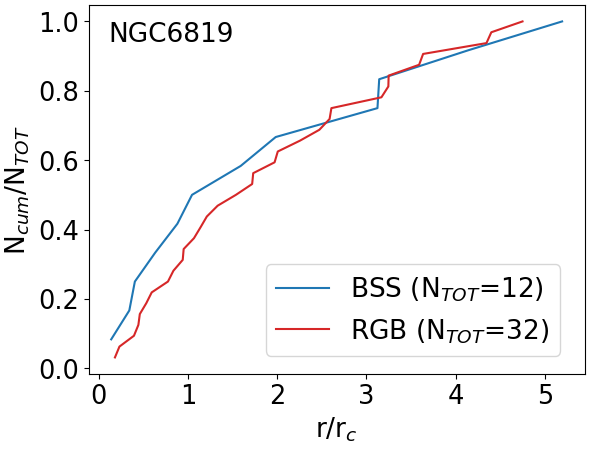}\par
\end{multicols}
\caption{The cumulative radial distributions of BSS and RGB populations. The normalized number of the two populations are plotted on the y-axis, and the radial distance in the units of core radius, $r_c$, is plotted on the x-axis.}
\label{Fig. 7}
\end{figure*} 

\subsubsection{Berkeley\,39} \label{Sec. 4.3.4}
We identify 23 BSS candidates in this cluster. \citet{al07} had reported 42 BSS in Berkeley 39. Only 16 of the AL07 BSS are identified as members in our criteria, of which 14 are our BSS candidates. We find 9 new BSS candidates in this cluster.  

\subsubsection{Melotte\,66} \label{Sec. 4.3.5}
Melotte 66 is the cluster with the least number of BSS in our analysis.  We found 14 BSS in this cluster. \citet{al07} had reported 35 BSS in Melotte 66, only 9 of their BSS are common with our BSS candidates. We find 5 new BSS candidates in this cluster.  

\subsubsection{NGC\,2158} \label{Sec. 4.3.6}
We found 40 BSS candidates in NGC 2158.  AL07 had also found 40 BSS candidates in this cluster, however, no positional information of these sources were found in the literature.   Hence, we could not make a comparison of our BSS candidates with AL07 BSS candidates.

\subsubsection{NGC\,2506} \label{Sec. 4.3.7}
We detect 28 BSS candidates in this cluster.  \citet{al07} listed 15 BSS in this cluster, 6 of which are common with our BSS candidates. We find 22 new BSS candidates, with two of them being bright BSS sources which have not been reported so far.

\subsection{BSS as Probes of Dynamical Evolution} \label{Sec. 4.4}
Being the most massive objects of the clusters, BSS are expected to be the most affected by the effect of dynamical friction that produces the segregation of massive stars toward the cluster center.  The bimodality in the BSS radial distributions was first discovered by \citet{ferraro97} in the globular cluster, M3, in their UV HST and ground-based optical wavelengths study of the cluster. Such a bimodality was then discovered in a large majority of the globular clusters, with only a small fraction of globular clusters showing no external upturn, and another small fraction showing completely flat BSS radial distributions, e.g., see references in \citet{ferraro14}. In their work, \citet{ferraro12} made use of RGB or horizontal branch stars (HB) as a reference population and plotted the ratio, $N_{\mathrm{BSS}}/N_{\mathrm{RGB}}$, against, the radial distance in units of $r_c$. The three families of clusters have undergone varying degrees of dynamical evolution with family I being the least dynamically evolved to family III being the most dynamically evolved. In family II clusters, the location of minima in the distribution, $r_\mathrm{{min}}$, systematically moves outward for more dynamically evolved clusters.  

We plotted cumulative radial distributions of BSS and RGB (our reference population) for these clusters. Figure~\ref{Fig. 7} shows the plots where normalized numbers of the two populations are plotted on the y-axis and the radial distance, in the units of the core radius ($r_c$), is plotted on the x-axis. For a meaningful comparison between the numbers of BSS and RGB, we consider only candidates within the same magnitude range of the two populations in this analysis (as mentioned on Figure~\ref{Fig. 7}).  In three clusters, Melotte 66, NGC 2158, and NGC 2506, BSS are more concentrated than the RGB for the entire extent of the cluster. Melotte 66 shows a dip in the BSS frequency distribution at $r$ $\sim$2.5$r_c$, but still shows BSS population to be more concentrated throughout the cluster extent. NGC 2506 shows no BSS in the regions beyond $r \sim$3$r_c$.  The radial distributions of three clusters, NGC 188, NGC 6791, and NGC 6819 show their BSS populations to be more concentrated in the inner regions, but less concentrated in the outer regions. In the cluster NGC 6819, the two distributions do not look much distinct in the outer regions, however in the central regions BSS population looks more concentrated. Berkeley 39 showed no differences in the radial distributions of the two populations.  For four of our clusters, the Kolmogorov-Smirnov test yields a high probability ($\geq$ 95\%) that the two populations, BSS and RGB, are not derived from the same parent population.  These are Melotte 66 (98.2\%), NGC 188 (96.8\%), NGC 2506 (99.8\%) and NGC 6791 (98.2\%). For NGC 2158, the test gives a low distinction, 88\% between the BSS and RGB populations.  

\begin{figure*}
\begin{multicols}{3}
\includegraphics[width=6.0cm]{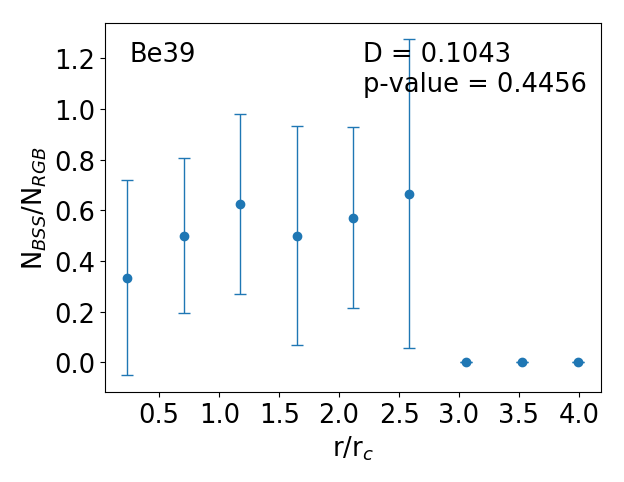}\par
\includegraphics[width=6.0cm]{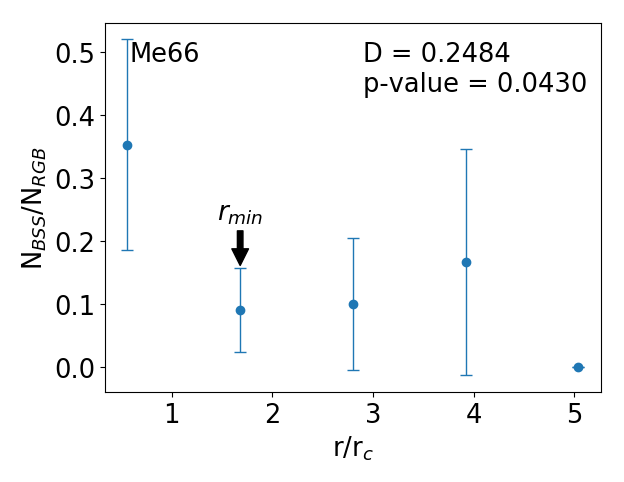}\par
\includegraphics[width=6.0cm]{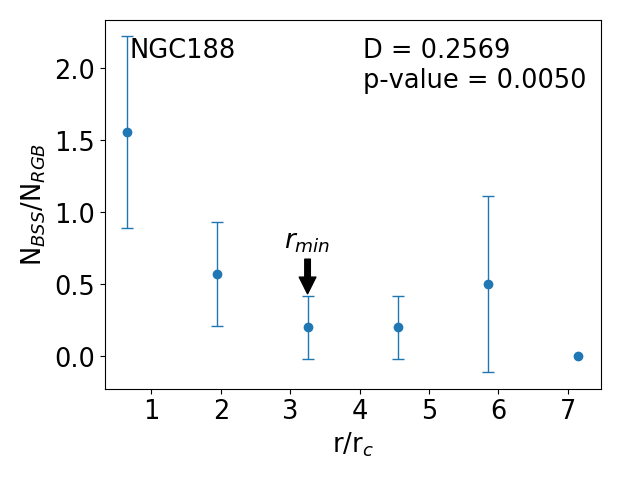}\par
\end{multicols}
\begin{multicols}{3}
\includegraphics[width=6.0cm]{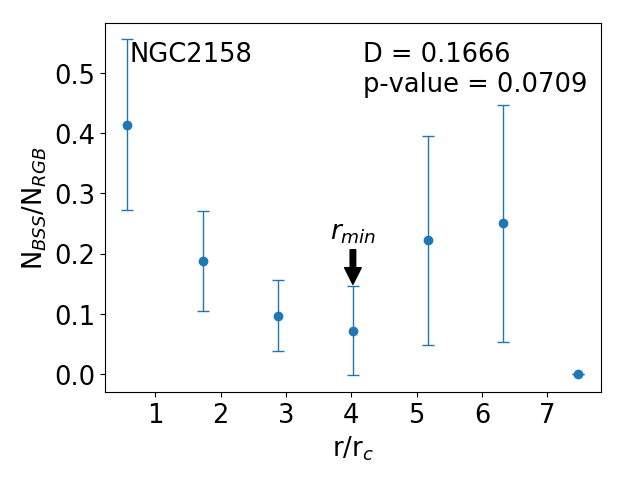}\par
\includegraphics[width=6.0cm]{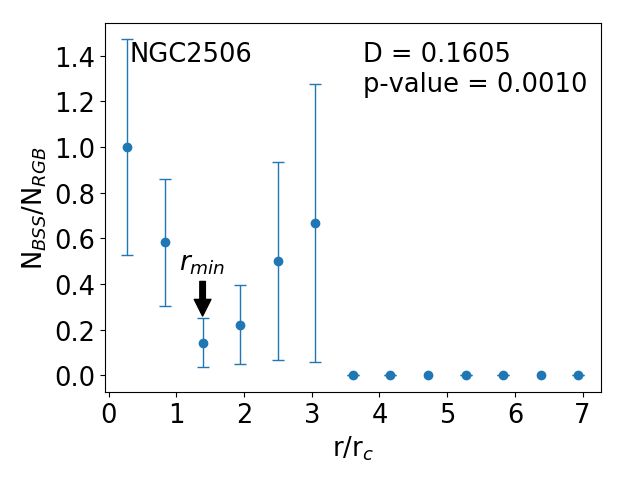}\par
\includegraphics[width=6.0cm]{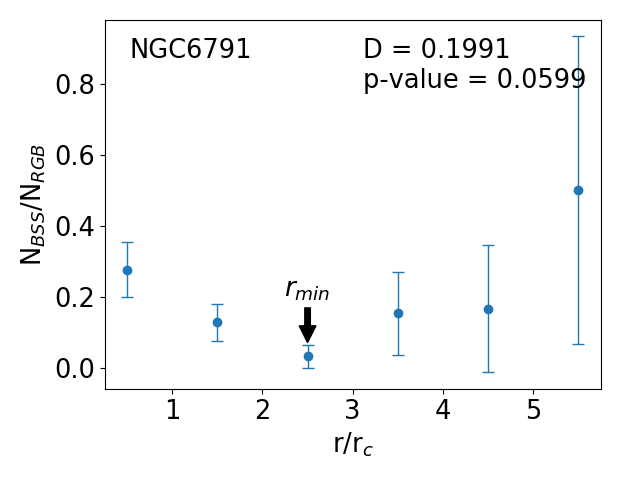}\par
\end{multicols}
\begin{multicols}{3}
\includegraphics[width=6.0cm]{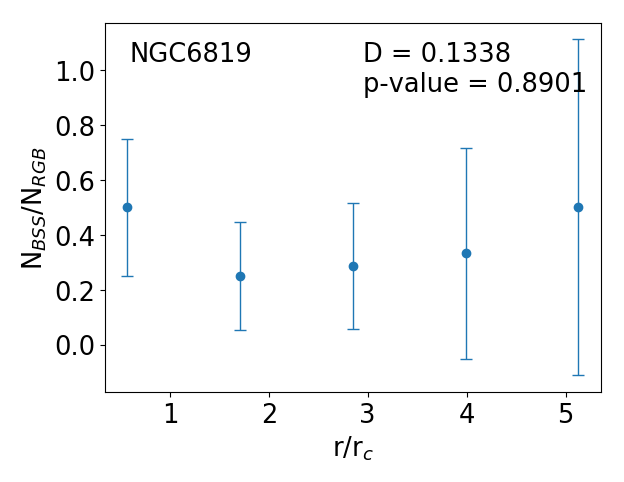}\par
\end{multicols}
\caption{The ratio $N_{\mathrm{BSS}}/N_{\mathrm{RGB}}$ is plotted against the radial distance in the units of core radius, $r_c$, for each cluster.  The bin sizes are selected such that all bins have at least 1 BSS except the end bins where there are no BSS left. Only the BSS and RGB in the same magnitude range for a given cluster are used for this analysis. The error bars represent Poisson errors.  The dip statistic, D, and the $p$-value from the dip test for bimodality are mentioned on the plots.}
\label{Fig. 8}
\end{figure*}

In Figure~\ref{Fig. 8}, we show the ratio of the numbers of the two populations, $N_{\mathrm{BSS}}/N_{\mathrm{RGB}}$, against the radial distance, $r$, in units of $r_c$.  As in Figure~\ref{Fig. 7}, in this analysis as well, we include only those BSS and RGB candidates which are in the same magnitude range in each cluster.  The $N_{\mathrm{BSS}}/N_{\mathrm{RGB}}$ are divided in equal sized bins in $r/r_c$ such that each bin has at least 1 BSS, except the end bins where there are no BSS left. In five of our clusters, Melotte 66, NGC 188, NGC 2158, NGC 2506, and NGC 6791, the radial distribution is seen to peak at the center, falling with radius until a certain radius, $r_\mathrm{{min}}$, and showing a rising trend again beyond this radius. The mean value of the bin in which the distribution falls to a minima, is called $r_\mathrm{{min}}$, and is listed in Table~\ref{table4} for each cluster. To evaluate the 
bimodality visible in these radial distributions, we performed Hartigan's dip statistic \citep{hartigan87} test.  The dip test, based on null-hypothesis logic, works by finding the maximum difference between the empirical distribution function and the unimodal distribution function that minimizes this maximum difference. The test result with $p$-values smaller than 0.05 suggests significant bimodality (or multimodality) in the distribution, and with $p$-values smaller than 0.1 but greater than 0.05 suggests marginal bimodality (or multimodality) in the distribution \citep{freeman12}.  The results of the dip test that we performed, the dip statistic (D) and the $p$-value, are mentioned on the plot of each cluster (Fig. \ref{Fig. 8}).
According to our analysis, Melotte 66, NGC 188, and NGC 2506 show significant bimodality whereas NGC 2158 and NGC 6791 show marginal bimodality.  We conclude that these five clusters are Family II type, as defined by \citet{ferraro12}, and are of intermediate dynamical ages. The remaining two clusters, Berkeley 39 and NGC 6819, show flat radial distributions. The dip tests for these two clusters also fail to detect bimodality in these two clusters. We thus classify these two clusters as Family I type, as defined in \citet{ferraro12}, but discuss their dynamical status in detail in Section~\ref{Sec. 5}.

\begin{figure}
\includegraphics[width=\columnwidth]{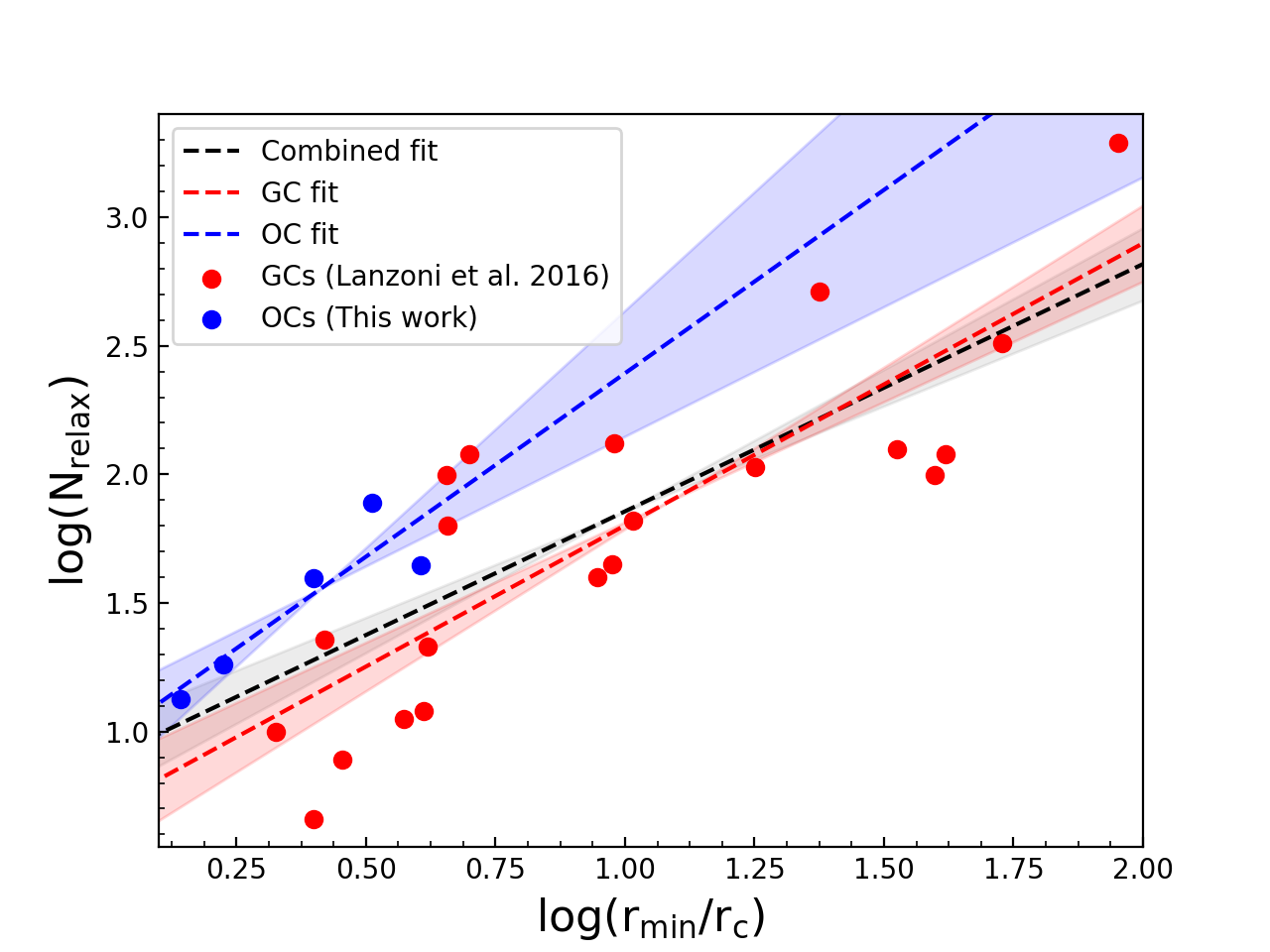}
\caption{The correlation between the location of minima in the BSS radial distributions, $r_{\mathrm{min}}/r_c$, and the number of current central relaxations occurred since cluster formation, $N_{\mathrm{relax}}$, for five open clusters (blue filled circles) and 21 globular clusters (red filled circles) of \citet{ferraro12} for which data has been taken from Table~1 of \citet{lanzoni16}.  The open cluster correlation shown as a blue dashed line, with errors calculated from the covariance matrix of the fit and shown as the blue shaded area, has a slope that is consistent within the errors, to the slope of the globular cluster correlation (red dashed line), but has a slightly larger intercept. The combined correlation, fitted to both the open and globular clusters, is shown as a black dashed line. The red and gray shaded areas show the errors of the globular cluster correlation, and the common correlation of open and globular clusters, respectively.}
\label{Fig. 9}
\end{figure} 

To test the scenario that the shift in the location of $r_{\mathrm{min}}$ in these five clusters with bimodal distributions, is suggestive of their dynamical ages, we estimated central relaxation times, $t_{rc}$, using the equation,  
$t_{rc}=1.491 \times 10^7 ~\mathrm{yr} \times \frac{k}{\mathrm{ln}(0.4N_{\ast})} \langle{m_{\ast}}\rangle^{-1} \rho_{M,0}^{1/2}~r_c^3$, where $k \approx$ 0.5592, $N_{\ast}$ is the estimated number of stars in the cluster, $\langle{m_{\ast}}\rangle$ is an average stellar mass in solar units, and $ \rho_{M,0}$ is the central mass density in $ M_{\sun}/pc^3 $ \citep{djorgovski93}. For this calculation, the average stellar mass and the central mass density were empirically estimated using the member stars for each cluster.  With these values of the central relaxation times, $t_{rc}$, and the cluster ages (Table~\ref{table4}), we estimated a theoretical parameter, indicative of the dynamical age of the cluster, $N_{\mathrm{relax}} = \mathrm{Age}/t_{rc}$, i.e. the numbers of current central relaxation times occurred since the cluster formation.  The two parameters, $t_{rc}$ and $N_{\mathrm{relax}}$, of the clusters are listed in Table~\ref{table4}. The plot between logarithm of $N_{\mathrm{relax}}$ and the logarithm of $r_{\mathrm{min}}/{r_c}$ for our Family II clusters (blue filled circles) is shown in Figure~\ref{Fig. 9}. On the same plot, we also show data-points of 21 globular clusters (red filled circles) taken from Table~1 of \citet{lanzoni16}. In order to plot the globular cluster data-points, we computed the $N_{\mathrm{relax}}$ values of the globular clusters by dividing the central relaxation times, $t_{rc}$, available in the Table~1 of \citet{lanzoni16} from the mean age of the globular clusters (=12 Gyr) \citep{forbes10} similar to that done by \citet{ferraro18}.  The best-fit relation for our open clusters data-points, shown as a blue dashed line, is:
\begin{equation}
\mathrm{log}(N_{\mathrm{relax}})= 1.43 (\pm 0.42)~\mathrm{log} (r_{\mathrm{min}}/r_c)+0.97 (\pm 0.17)
\label{eq1}
\end{equation}

The errors are calculated from the covariance matrix of the fit and are used to plot the shaded blue region.  The correlation confirms that the observed parameter, $r_{\mathrm{min}}$, does reveal the true dynamical status of the clusters.  
Interestingly, the slope of the fitted open cluster correlation turns out to be in agreement, within the errors, to the slope of the previously known correlation of globular clusters \citep{ferraro12}\footnote{\footnotesize{\citet{ferraro12} reported a correlation between $log(t_{rc}/t_{\mathrm{H}})$ and $r_{\mathrm{min}}/{r_c}$, where $t_{\mathrm{H}}$ is the Hubble time, for 21 coeval globular clusters that are included in \citet{lanzoni16}, as: $\mathrm{log}(t_{rc}/t_H)=-1.11\mathrm{log}(r_{\mathrm{min}}/r_c)-0.78$.  \citet{ferraro12} noted that using individual globular cluster ages in place of the Hubble time, does not change the correlation.  As we are fitting here $N_{\mathrm{relax}}$ for these 21 globular clusters, our values of the slopes and intercepts of the globular cluster correlation are positive.}}, shown as red dashed line:
\begin{equation}
\mathrm{log}(N_{\mathrm{relax}})= 1.09 (\pm 0.16)~\mathrm{log}(r_{\mathrm{min}}/r_c)+0.7 (\pm 0.17)
\end{equation}
As our open cluster data-points share a common parameter space as the globular cluster data-points, we attempt a combined fit for both the clusters, and the correlation (black dashed line) is given as:
\begin{equation}
\mathrm{log}(N_{\mathrm{relax}})= 0.96 (\pm 0.14)~\mathrm{log}(r_{\mathrm{min}}/r_c)+0.89 (\pm 0.14) 
\end{equation}
 
The open cluster correlation equation \ref{eq1} has a higher intercept.  This could be due to the structural differences in the two kinds of clusters which give rise to longer relaxation times for globular clusters, or could be a signature of other factors, such as dominance of different formation channels of BSS in the two kinds of clusters (discussed further in Sec. \ref{Sec. 5}). In order to obtain a more constrained correlation for open clusters, we certainly need to analyze more open clusters with reasonable numbers of BSS (In preparation, Rao et al. 2020).

\section{Discussion} \label{Sec. 5}
The Gaia DR2 data has made it feasible to identify secure cluster members using the precise proper motions, parallaxes, and radial velocity information that it provides. We evolve a criteria of membership determination by the combined use of all these parameters along with the information of confirmed members of the clusters found in the literature.  Our membership criteria has been fixed by using various selection ranges in proper motions and parallaxes, and after arriving at a compromise between minimizing the visible contamination on the CMDs and maximizing the retrieval of known members of the clusters by our criteria.  The mean of the proper motions of our members, determined by fitting the Gaussian functions as well as by employing the mean shift-based algorithm, match very well with the mean proper motions of the known members of the clusters.  The mean distances of our bright (G $\leq$15 mag) members are in excellent agreement with the mean distances of the known members of the clusters.   In five of our clusters, we retrieve 90--100\%, and in the remaining two clusters, 80--90\% of all previously known members, most of which are RGB stars, but in some clusters they also include main-sequence stars below the turn-off point.  Our membership criteria maybe stringent as we do miss a small fraction of the confirmed giants and BSS stars, however, since our analysis is based on the BSS and the reference population (RGB) having the same magnitude range, our incompleteness in both the populations is expected to be comparable and hence irrelevant to the analysis presented here.  

There are large inconsistencies in AL07 BSS candidates.  From our matched clusters, 40--60\% of the AL07 BSS are found non-members. Thus a membership analysis of cluster members using kinematic information, such as that provided by Gaia DR2, is  very important to study BSS populations of open clusters.  In the clusters with spectroscopically known BSS, NGC 188, NGC 6791 and NGC 6819, we retrieve 70--100\% of the BSS but lose up to 30\% BSS due to our stringent kinematic selection criteria.  At the same time, we find 8 new BSS candidates in NGC 188, 30 new BSS candidates in NGC 6791 and 2 new BSS candidates in NGC 6819. Among our new BSS candidates of these three clusters, the fraction of non-members are 12\% in NGC 6791 and 20\% in NGC 188, and a significant 66\% in NGC 6819.  The crowdedness along the line of sight for NGC 6819 (Figure~\ref{Fig. A3}) explains the high field star fraction in case of this cluster. A radial velocity follow-up of the identified BSS candidates is necessary for membership confirmation. 

BSS radial distributions can be exploited to learn about the dynamical ages of the clusters. The common feature of bimodal BSS radial distributions in globular clusters \citep{ferraro12, beccari13} has been reproduced using the numerical simulations by \citet{mapelli04, mapelli06} and \citet{lanzoni07}.  These numerical simulations have shown that the central peak in the radial distribution can be attributed to both, formation of collisional BSS in the densest regions of the cluster and to sinking of mass-transfer binaries in the central regions as a result of mass segregation.  The secondary maxima occurring in the outskirts of the clusters, on the other hand, is attributed to BSS generated by primordial binaries via the mass transfer process in a largely non-interactive manner from other cluster stars.  Whereas this observational signature of mass segregation has long been recognized and extensively exploited as an accurate dynamical clock in globular clusters, we present the first similar analysis for multiple open clusters and notice an extension
of such a correlation in open clusters.       

NGC 2158 is the most dynamically evolved cluster as per our analysis.  Its short relaxation time corroborates our finding that the effect of dynamical friction has significantly altered the initial distribution of its most massive population up to more than half of its radius ($r_{\mathrm{min}}=5.5\arcmin$).  NGC 2158 has been known to be an interesting cluster possessing a globular cluster like core \citep{arp62}, low metallicity, [Fe/H]=$-$0.64, a young age of 1--2 Gyr \citep{carraro02}, and location in the Galactic plane.  Our result provides first direct evidence of the cluster having an old dynamical age despite being a relatively young open cluster (1.9 Gyr), among the youngest we study in this work.

We find that NGC 188 is the second most dynamically evolved cluster.  Even though the dip in the minima in NGC 188 occurs at a smaller core radius (3.25$r_c$) as compared to its location in NGC 2158 (4.0$r_c$), NGC 188 has the highest central peak in the distribution, higher by a factor of 3--5 times from most clusters and higher by a factor of 1--2 times from NGC 2506.  Our finding is consistent with the previous knowledge that NGC 188 is a dynamically evolved cluster as shown by \citet{geller13} based on N-body simulations of the cluster alongside the observational data of the cluster. Melotte 66 and NGC 2506 appear of comparable dynamical age from their BSS radial distribution as is also confirmed by their relaxation times. 

Based on their flat radial distributions, Berkeley 39 and NGC 6819 may be classified as Family I type clusters.  The theoretically estimated values of $N_{\mathrm{relax}}$ for these two clusters (see Table~\ref{table4}), however, suggest that these two clusters are dynamically evolved.  With these $N_{\mathrm{relax}}$ values of Berkeley 39 and NGC 6819, 
their $r_{\mathrm{min}}$ are predicted to be 4.36$r_c$ and 2.75$r_c$, respectively, following the open cluster correlation 
(Eq. \ref{eq1}). In Berkeley 39, however,   
all the BSS of the cluster can be seen to be already segregated to cluster regions within $r \sim$3$r_c$, i.e. smaller than the predicted $r_{\mathrm{min}}$. This indicates that Berkeley 39 has undergone mass segregation and is dynamically evolved, and could in fact be a Family III cluster.  The RGB stars of this cluster too are segregated to inner cluster regions with radial distances smaller than $\sim$4$r_c$.  With the segregation of two of its most massive stellar populations complete to regions within the inner half of the cluster radius, i.e. $\sim$7.36$r_c$ (or 14$\arcmin$), this cluster is indeed dynamically evolved. Without a clear central peak in the BSS radial distribution though, Berkeley 39 cannot be definitively classified as a Family III cluster. In the cluster NGC 6819, we do not see a minima in the BSS radial distribution at the predicted $r_{\mathrm{min}} \sim 2.75 r_c$ of the cluster.  Thus, with the current data that we use, we are unable to comment on whether NGC 6819 is likely a Family II cluster or a Family I cluster as its radial distribution shows.  With radial velocity information for our BSS and RGB candidates, it may be possible to revisit the Family classification of Berkeley 39 and NGC 6819.

The location of minima in BSS radial distributions of globular clusters has been shown to work as the ``hand'' of the dynamical clock by \citet{ferraro12}. For the first time in the literature, we investigate whether BSS radial distributions of open clusters do also work as accurate probes of dynamical ages of clusters. Though we have only five open clusters with bimodal BSS distributions, we clearly see a positive correlation between $r_{\mathrm{min}}/r_c$ and the dynamical ages of the clusters, quantified here by $N_{\mathrm{Relax}}$, the number of current central relaxations experienced by clusters during their age. This correlation confirms that BSS radial distributions are sensitive probes of dynamical evolution in open clusters as well.

Our best-fit open cluster correlation is comparable to the previously known globular cluster correlation in its slope, but has a higher intercept.  A possible reason for an offset in the two intercepts could be pertaining to the structural differences among the two kinds of clusters.  Apart from the vast differences in the numbers of stars, metallicity, and stellar density in these two kinds of clusters, there are also large variations in their sizes. For example, the most evolved Family II cluster (Radius $\sim$8$r_c$) in our sample has its $r_{\mathrm{min}}$ equal to 4.0$r_c$.  Indeed as can be seen in \citet{kharchenko13}, the radii of most open clusters are typically smaller than ten times the King's core radii of these clusters. An implication of the small sizes of open clusters is that, the range of $r_{\mathrm{min}}$ in Family II open clusters will vary between 0 and $\sim$10$r_c$. Clusters with $r_{\mathrm{min}}$ greater than $\sim$10$r_c$ would have already transitioned to Family III type.  This is unlike Family II globular clusters which show a much larger range of $r_{\mathrm{min}}$ (0--100$r_c$; see Fig. \ref{Fig. 9}). The offset in the two intercepts might also hint at the dominance of different formation channel in the two kinds of clusters.  As shown by \citet[and the references therein]{chat13} stellar collisions are shown to contribute to BSS formation in dense cluster environments such as in globular clusters and in the cores of open clusters. However, collisions are not considered to be the dominant formation channel in open clusters \citep{leonard96,hurly05,perets09}. Most notably, in two open clusters, M67 and NGC 188, radial velocity surveys have shown a much higher binary fraction in BSS, 60$\pm$24\% and 76$\pm$22\% respectively \citep{latham07,geller08}. 
We stress that extending similar analysis to a larger sample of open clusters containing reasonable number of BSS populations, as well as confirmation of membership of BSS candidates by new radial velocity observations, would be important steps in further improving our knowledge of this correlation in open clusters. Such a correlation, if established, would serve as the basis for the use of BSS radial distributions as direct indicators of the dynamical evolution of open clusters.

\section*{Acknowledgements}
The authors are grateful to the anonymous referee for the valuable comments.  SB acknowledges support from the IMPRS on Astrophysics at the LMU Munich.  This work has made use of data from the European Space Agency (ESA) mission
{\it Gaia} (\url{https://www.cosmos.esa.int/gaia}), processed by the {\it Gaia}
Data Processing and Analysis Consortium (DPAC, \url{https://www.cosmos.esa.int/web/gaia/dpac/consortium}). Funding for the DPAC has been provided by national institutions, in particular the institutions participating in the {\it Gaia} Multilateral Agreement. This research has made use of the VizieR catalogue access tool, CDS, Strasbourg, France. This research made use of Astropy-- a community-developed core Python package for Astronomy \citep{2013A&A...558A..33A}, Numpy \citep{oliphant15} and Matplotlib \citep{hunter07}. This research also made use of NASA's Astrophysics Data System (ADS).

\appendix
\section{Additional Figures}
\begin{figure*}[htb]
\adjustbox{max width=\textwidth}{\includegraphics[width=17.5cm]{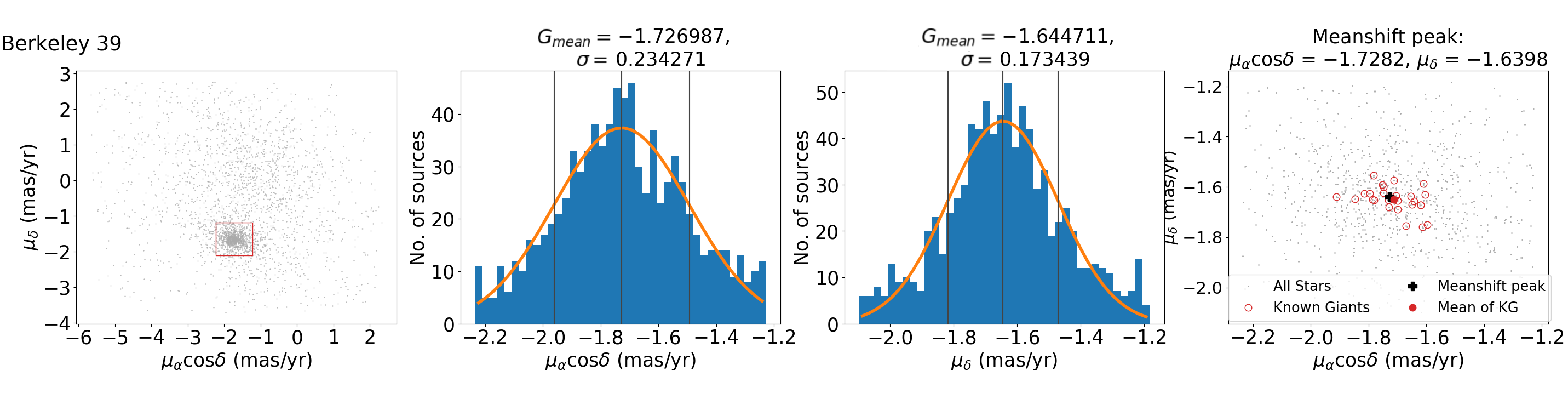}}
\adjustbox{max width=\textwidth}{\includegraphics[width=17.5cm]{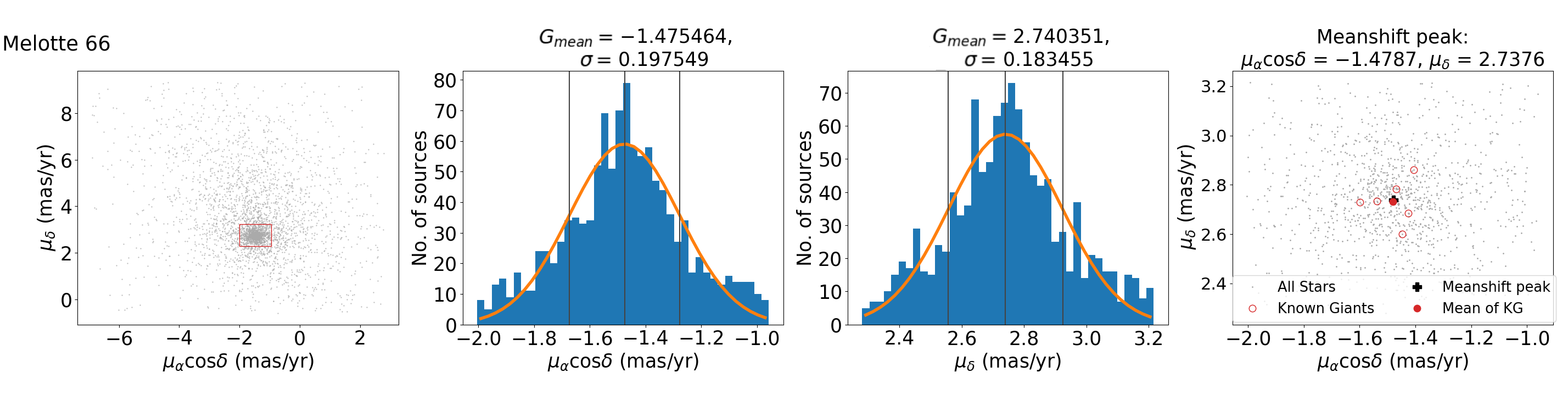}}
\adjustbox{max width=\textwidth}{\includegraphics[width=17.5cm]{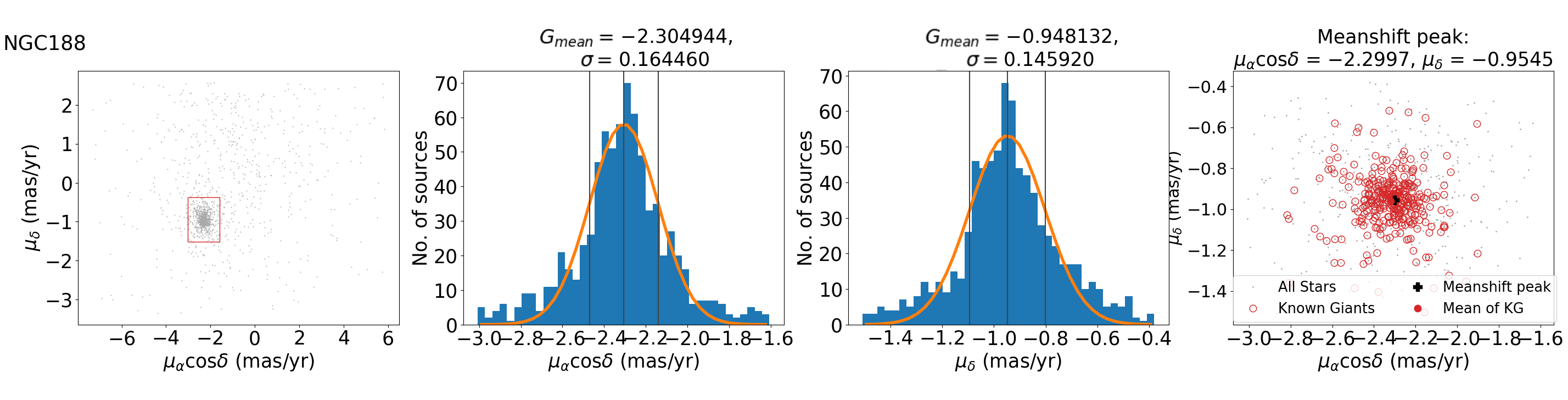}}
\end{figure*}

%\clearpage

\begin{figure*}
\adjustbox{max width=\textwidth}{\includegraphics[width=17.5cm]{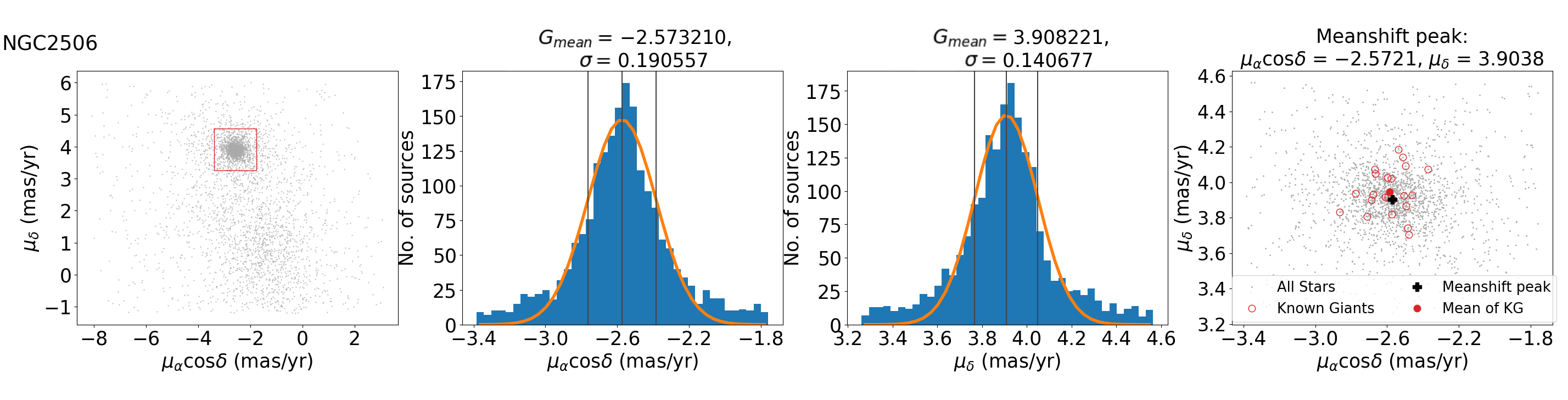}}
\adjustbox{max width=\textwidth}{\includegraphics[width=17.5cm]{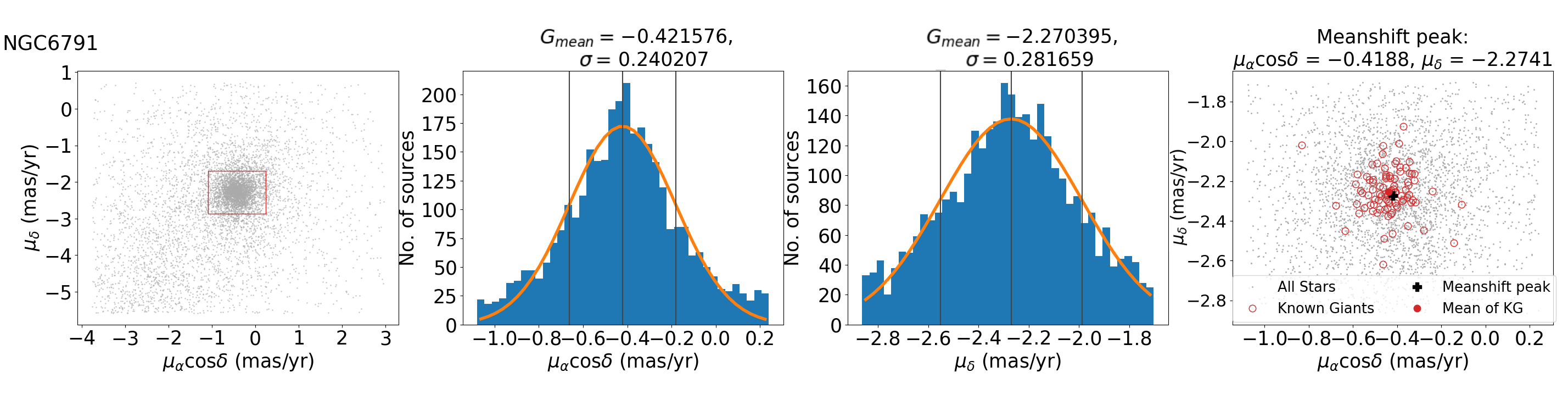}}
\adjustbox{max width=\textwidth}{\includegraphics[width=17.5cm]{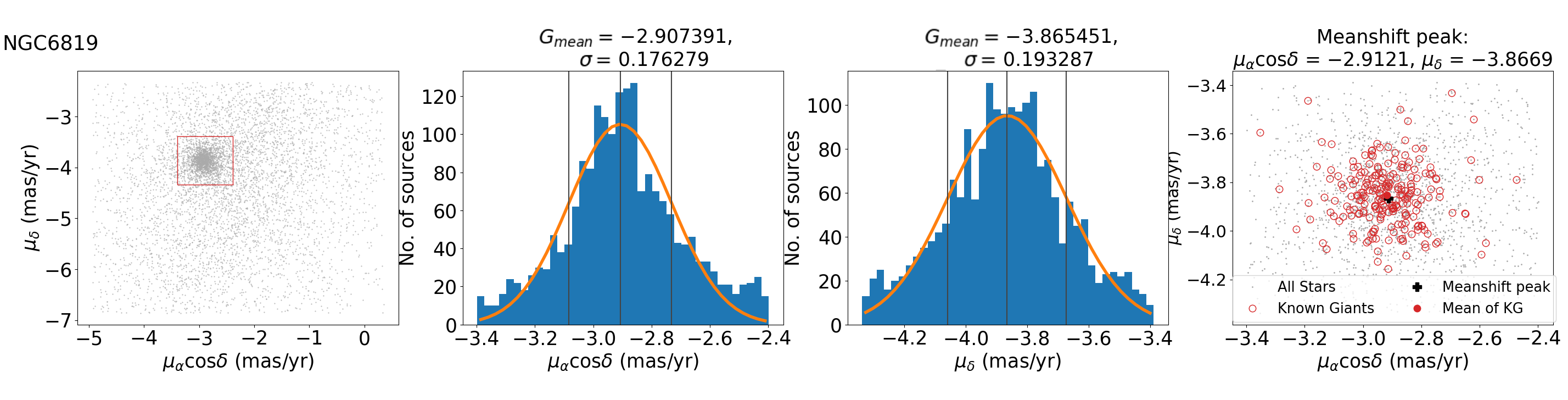}}
    \caption{The left panel shows a scatter diagram of proper motions of sources within 10$\arcmin$.  The rectangular region on this panel shows our initial range of proper motions fixed by visual examination of the distributions.  The two middle panels show the histograms of proper motions in RA and DEC of the selected sources, respectively. The fitted Gaussian functions are overplotted on the distributions. The right panels show the distribution of proper motions of selected sources, with the proper motions of previously confirmed spectroscopic members shown as red open circles, and their mean position shown with a red filled circle. The peak position calculated from the mean shift algorithm is shown as a black plus symbol.}
    \label{Fig. A1}
    \end{figure*} 
\clearpage

\begin{figure*}
\begin{multicols}{2}
\includegraphics[width=9cm]{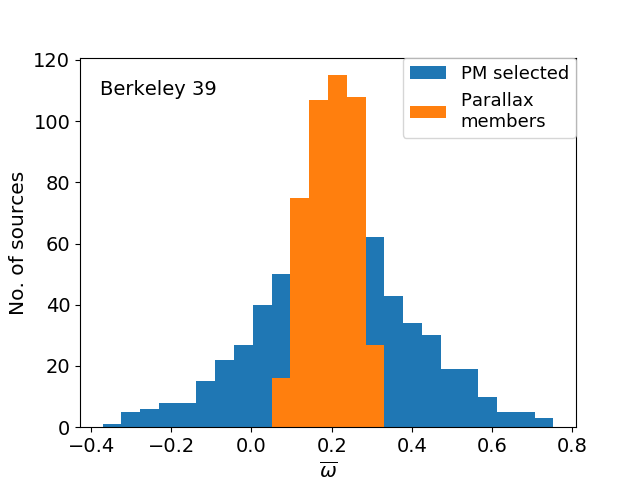}\par 
\includegraphics[width=9cm]{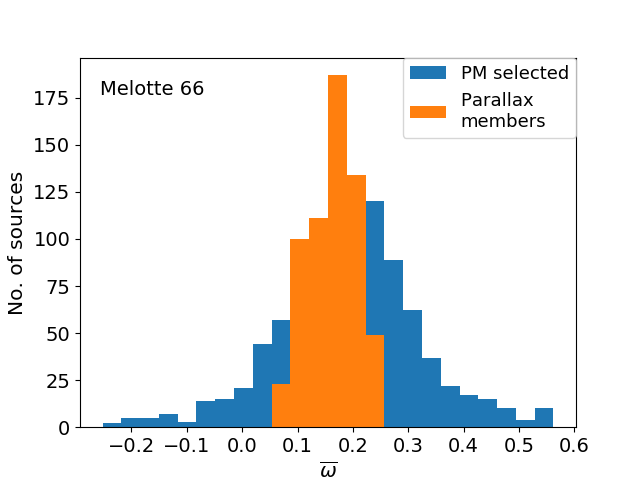}\par
\end{multicols}
    \begin{multicols}{2}
\includegraphics[width=9cm]{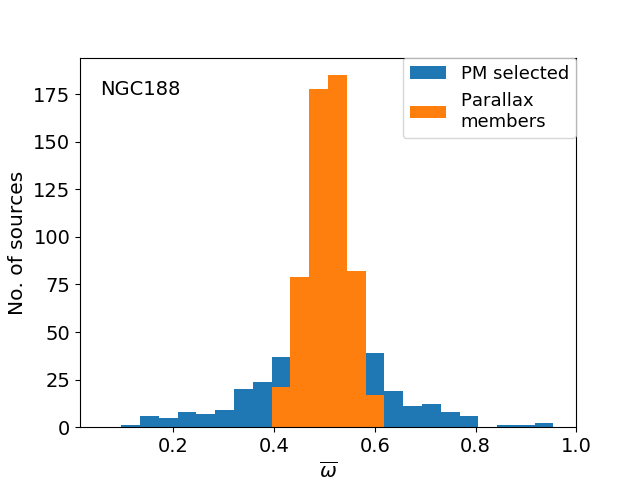}\par
\includegraphics[width=9cm]{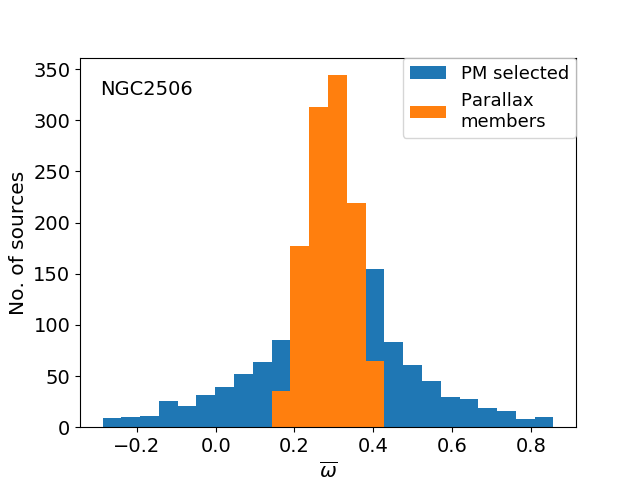}\par
\end{multicols}
 \begin{multicols}{2}
\includegraphics[width=9cm]{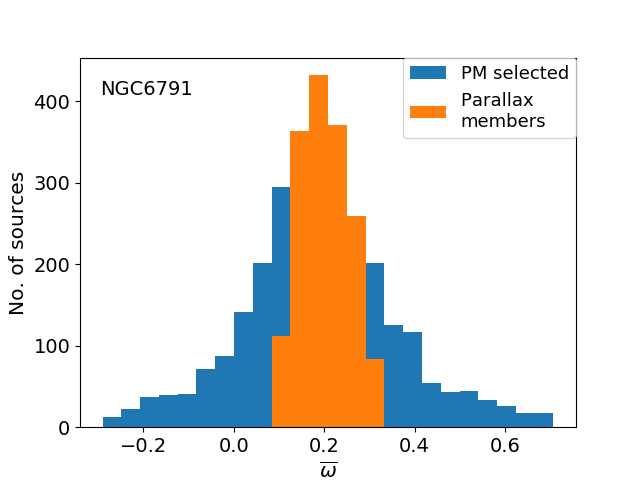}\par
\includegraphics[width=9cm]{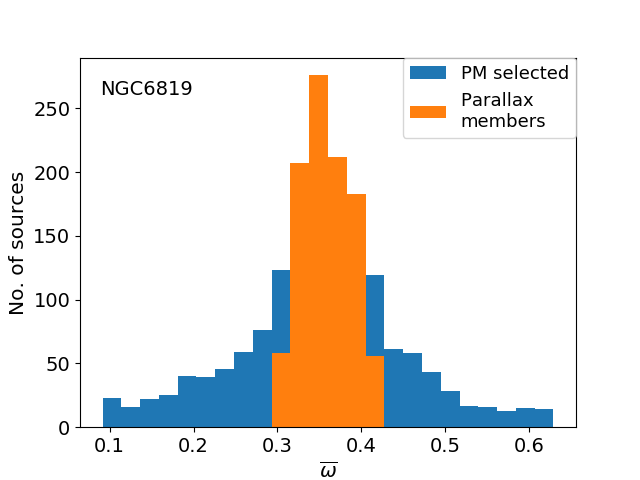}\par
\end{multicols}
\caption{The parallax distribution of proper motion selected sources is shown in blue histogram.  The orange histogram shows the parallax distribution of proper motion selected sources with their parallaxes within the range $\bar{\omega}$ $\pm 3 \Delta \bar{\omega}$, where $\bar{\omega}$ is the mean Gaia DR2 parallax, and $\Delta \bar{\omega}$ is the mean error in the Gaia DR2 parallaxes of the previously known, spectroscopically confirmed members of the clusters.}
\label{Fig. A2}
\end{figure*} 

\begin{figure*}
\begin{multicols}{2}
\includegraphics[width=9cm]{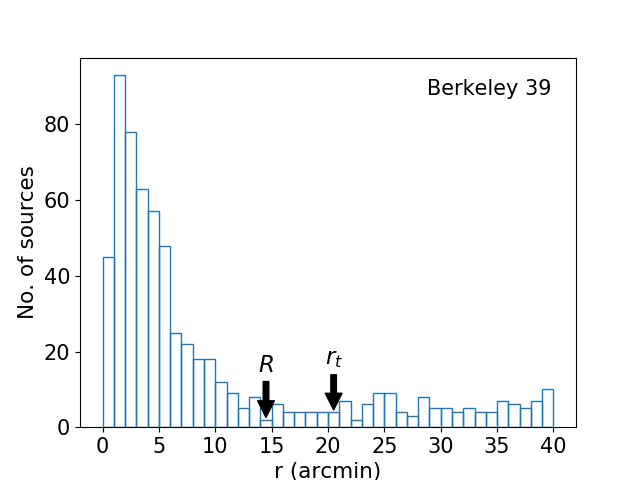}\par
\includegraphics[width=9cm]{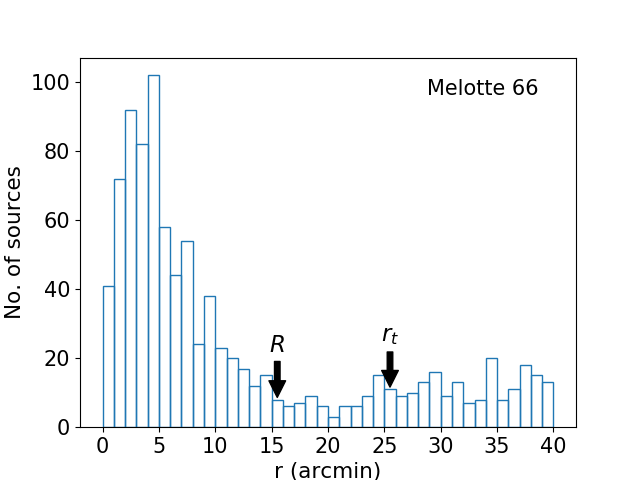}\par
\end{multicols}
\begin{multicols}{2}
\includegraphics[width=9cm]{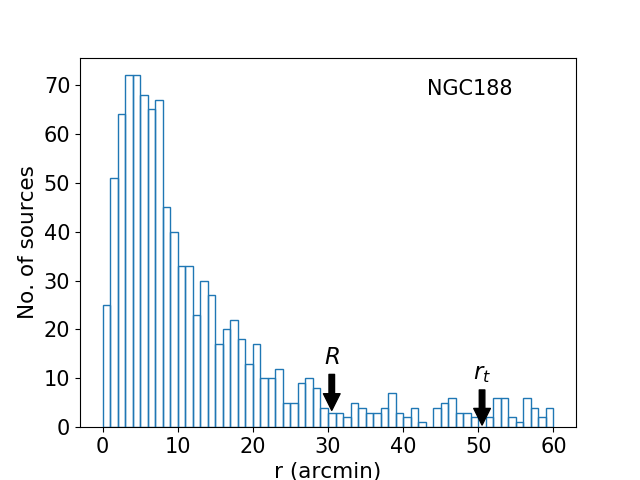}\par
\includegraphics[width=9cm]{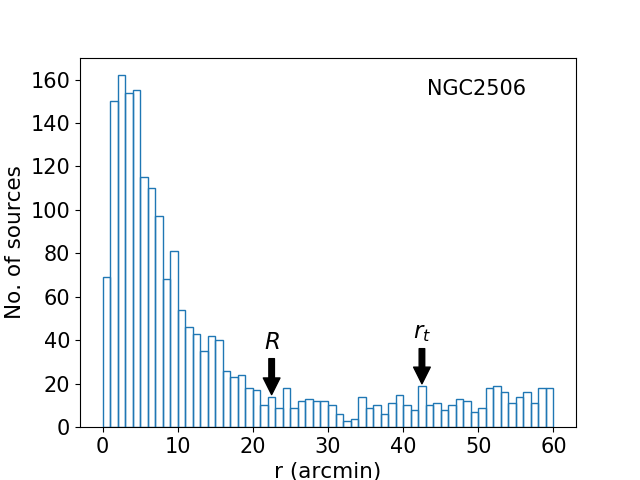}\par
\end{multicols}
\begin{multicols}{2}
\includegraphics[width=9cm]{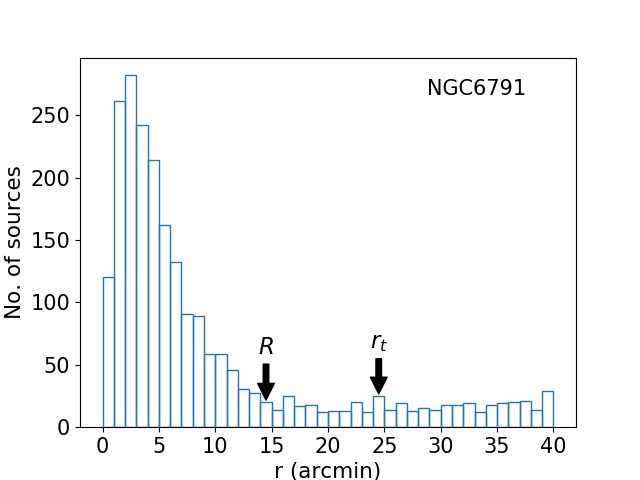}\par
\includegraphics[width=9cm]{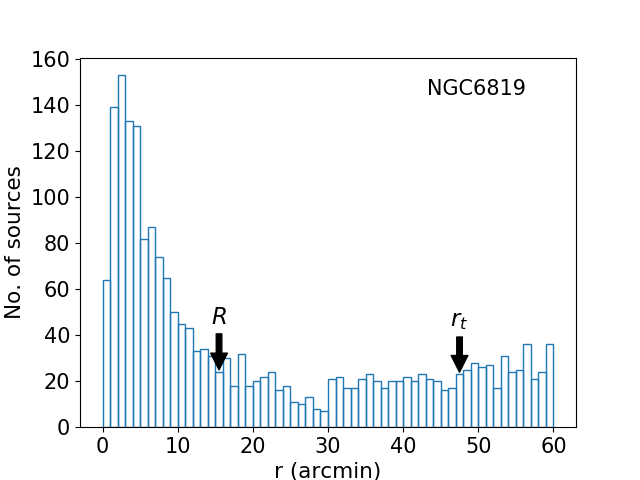}\par
\end{multicols}
 \caption{The radial distributions of sources following the proper motion and parallax ranges of selection of cluster members.  The estimated cluster radii (R) and tidal radii ($r_t$) are marked on the figures.}
\label{Fig. A3}
 \end{figure*} 
\clearpage

\begin{figure*}
    \includegraphics[width=17.5cm]{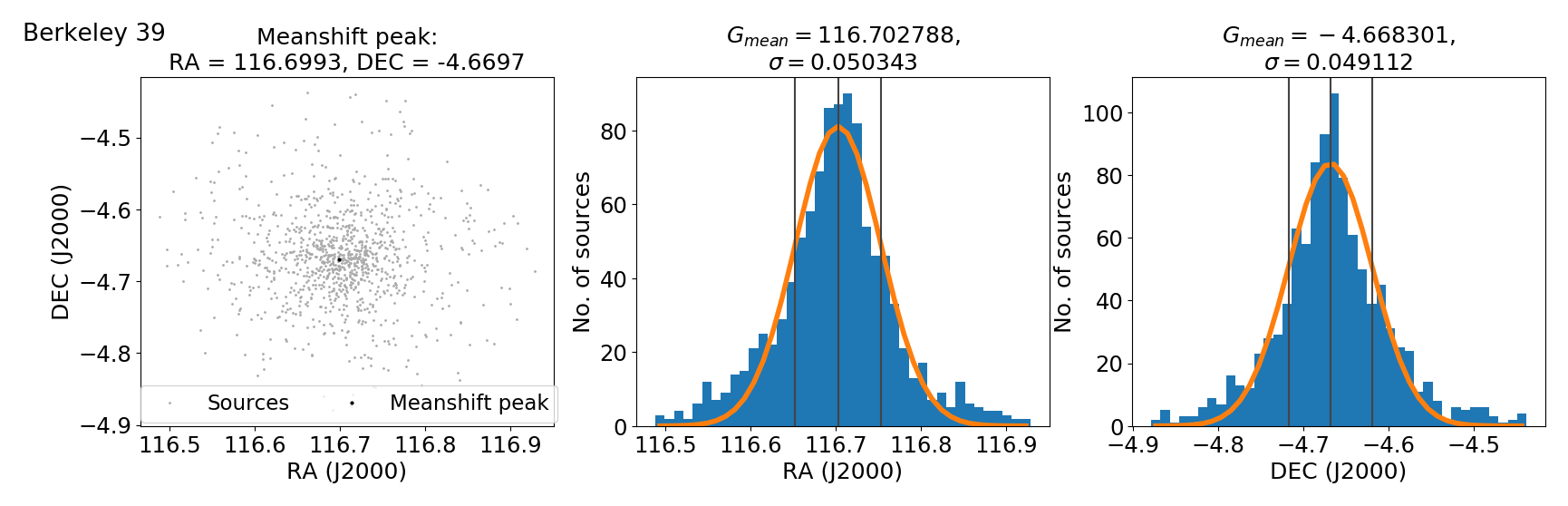}\par
    \includegraphics[width=17.5cm]{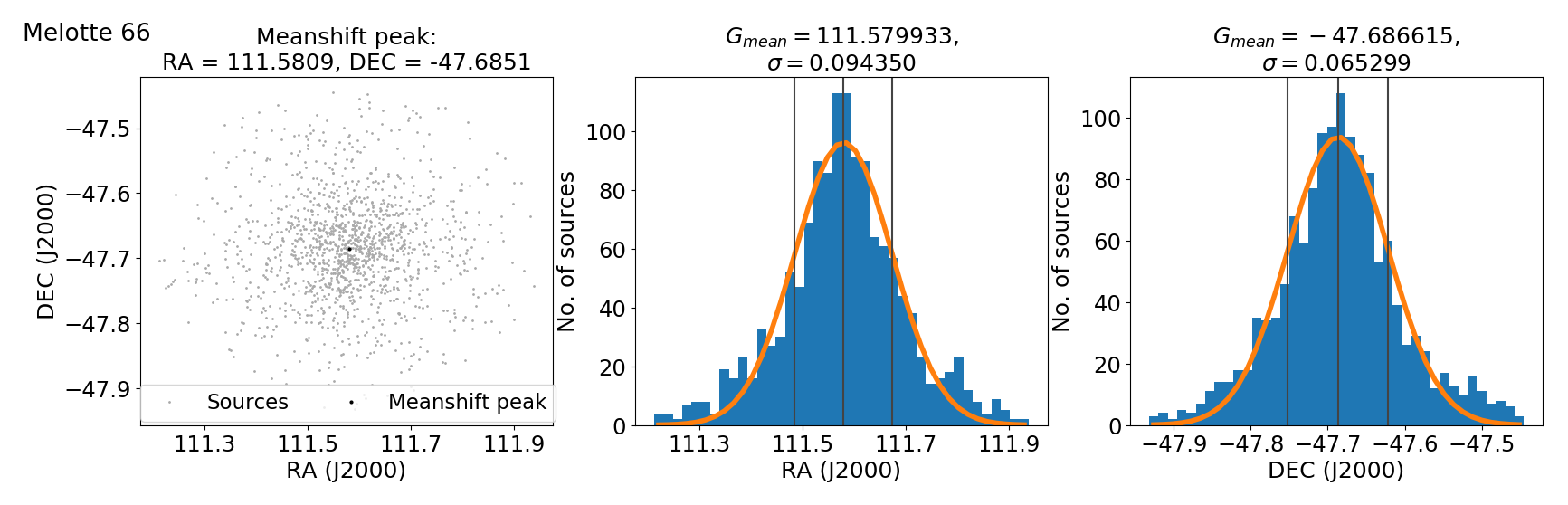}\par
    \includegraphics[width=17.5cm]{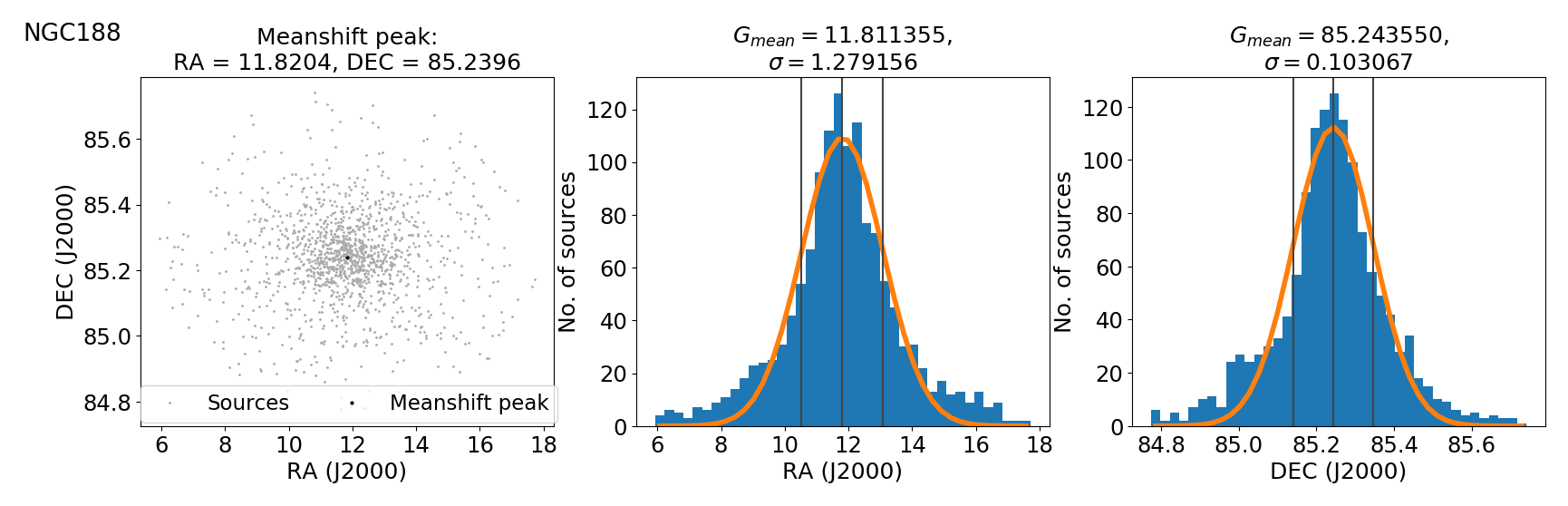}\par
\end{figure*}

    \clearpage   
    \begin{figure*}
    \includegraphics[width=17.5cm]{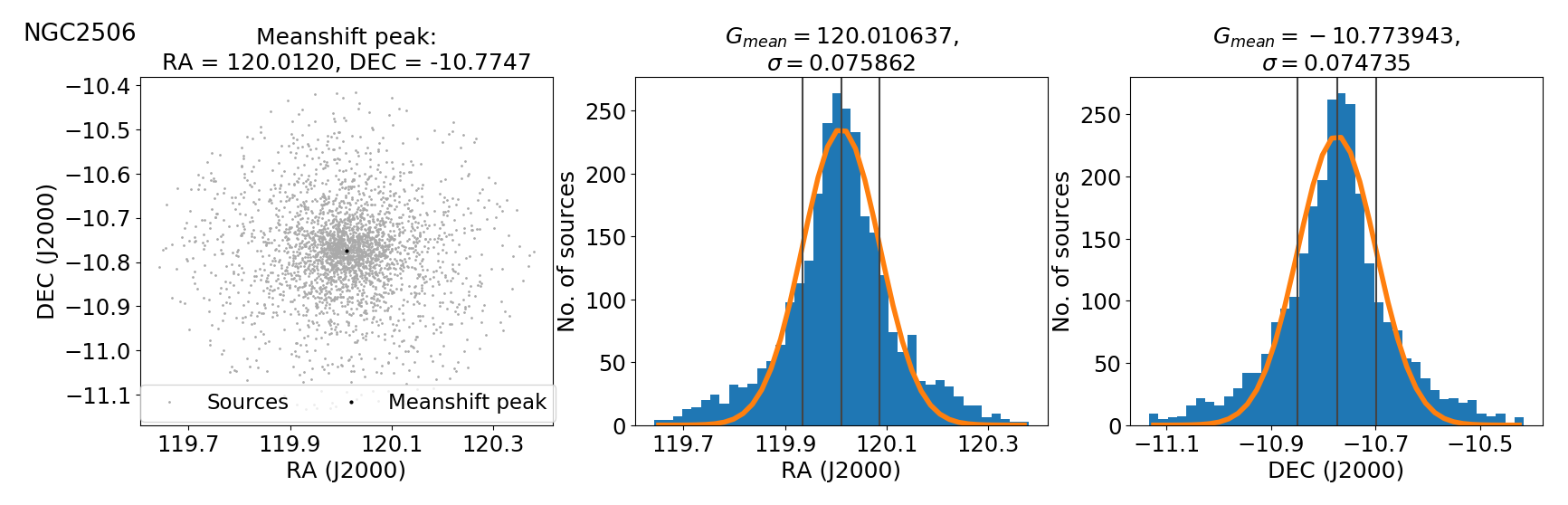}\par
    \includegraphics[width=17.5cm]{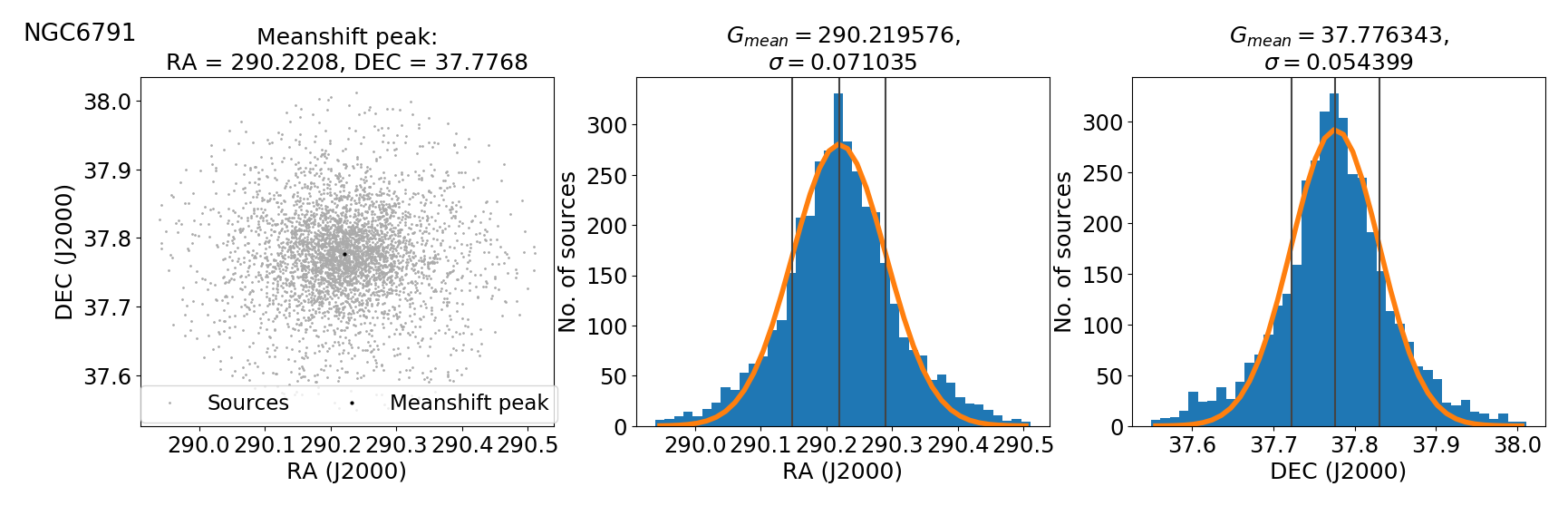}\par
    \includegraphics[width=17.5cm]{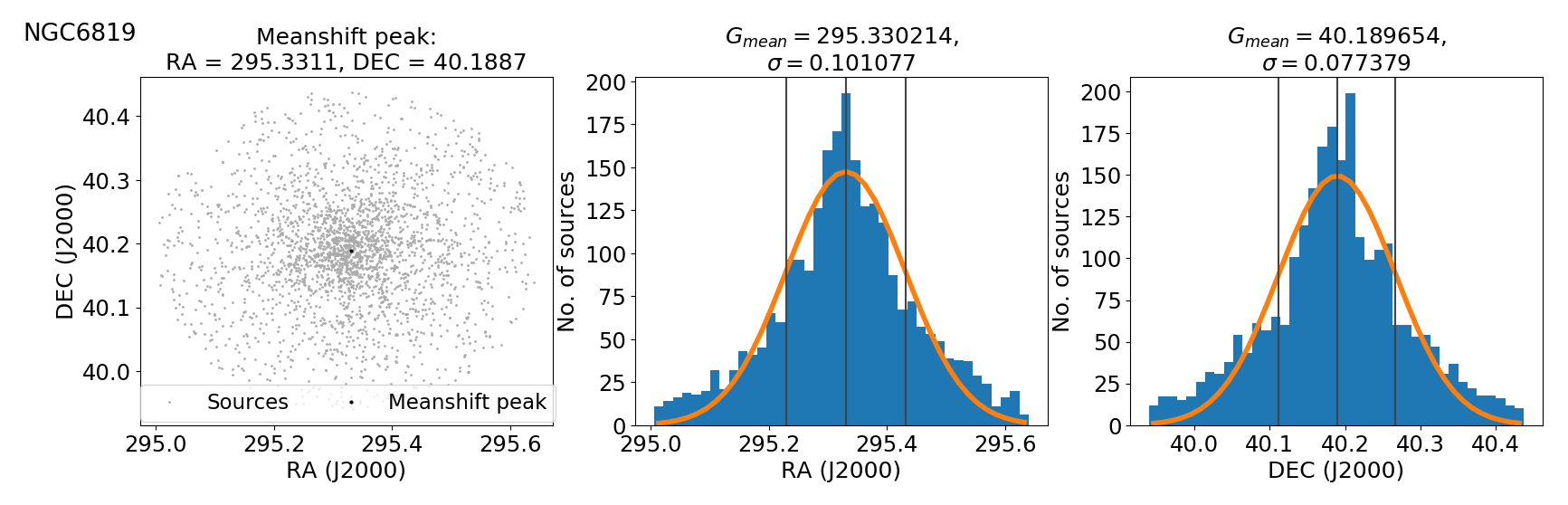}\par
    \caption{The left panels show the result of the mean shift clustering algorithm to determine the cluster center.  The middle and the right panels show the frequency distributions of cluster members in RA and DEC, and a Gaussian function fitted to the distribution.}
    \label{Fig. A4}
    \end{figure*} 
\clearpage
\begin{figure*}
\begin{multicols}{2}
\includegraphics[width=9cm]{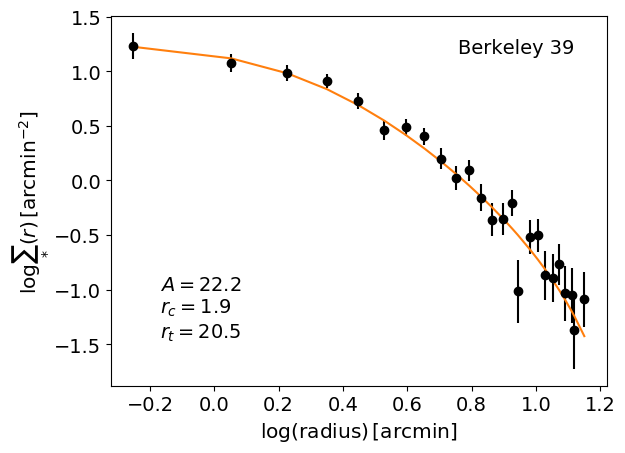}\par
\includegraphics[width=9cm]{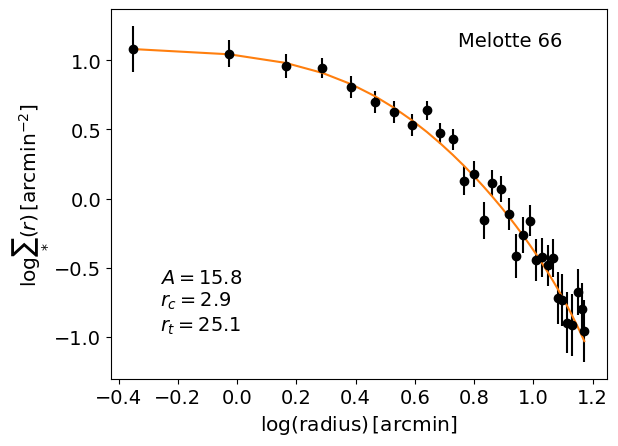}\par
\end{multicols}
\begin{multicols}{2}
\includegraphics[width=9cm]{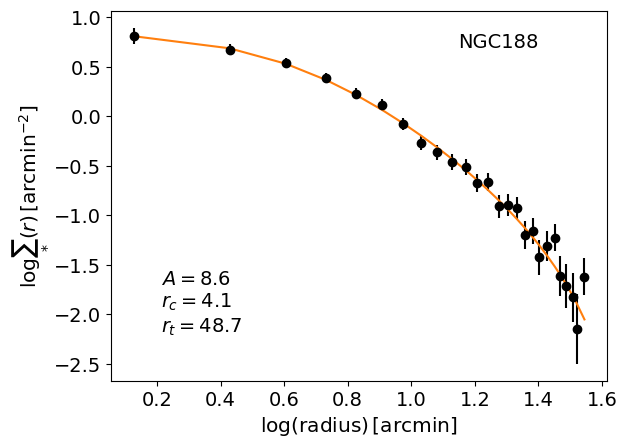}\par
\includegraphics[width=9cm]{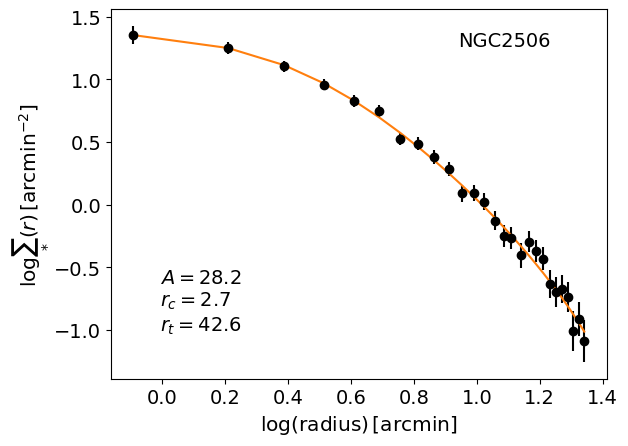}\par
\end{multicols}
\begin{multicols}{2}
\includegraphics[width=9cm]{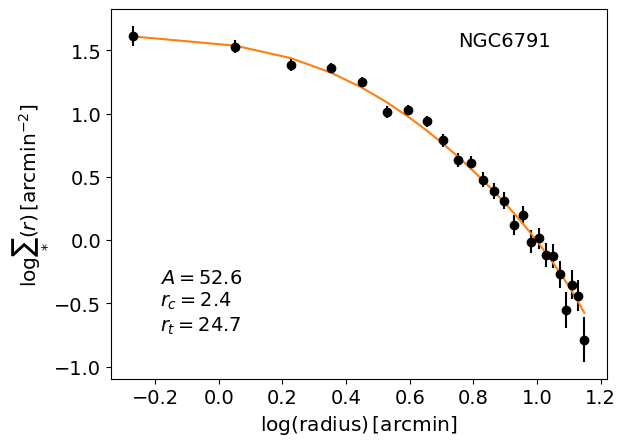}\par
\includegraphics[width=9cm]{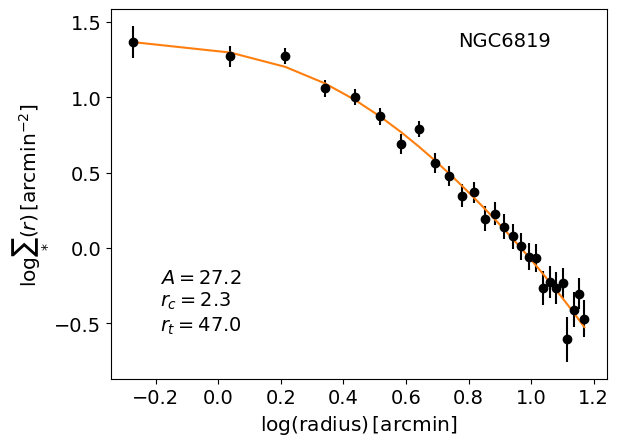}\par
\end{multicols}
 \caption{The radial density profile of the cluster members are shown with King's function fitting. The error bars represent 1$\sigma$ Poisson errors.}
\label{Fig. A5}
 \end{figure*}

 \begin{figure*}
\begin{multicols}{2}
\includegraphics[width=10cm]{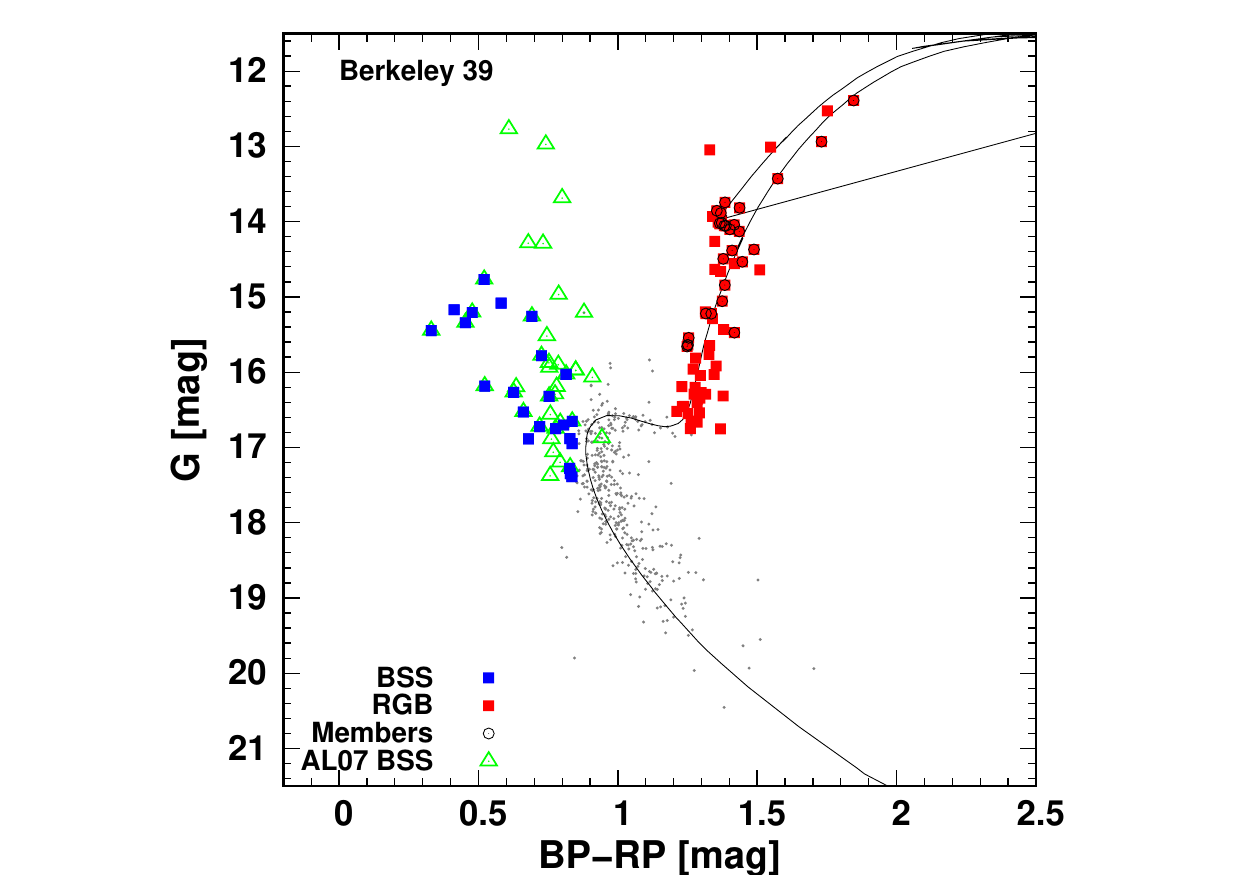}\par
\includegraphics[width=10cm]{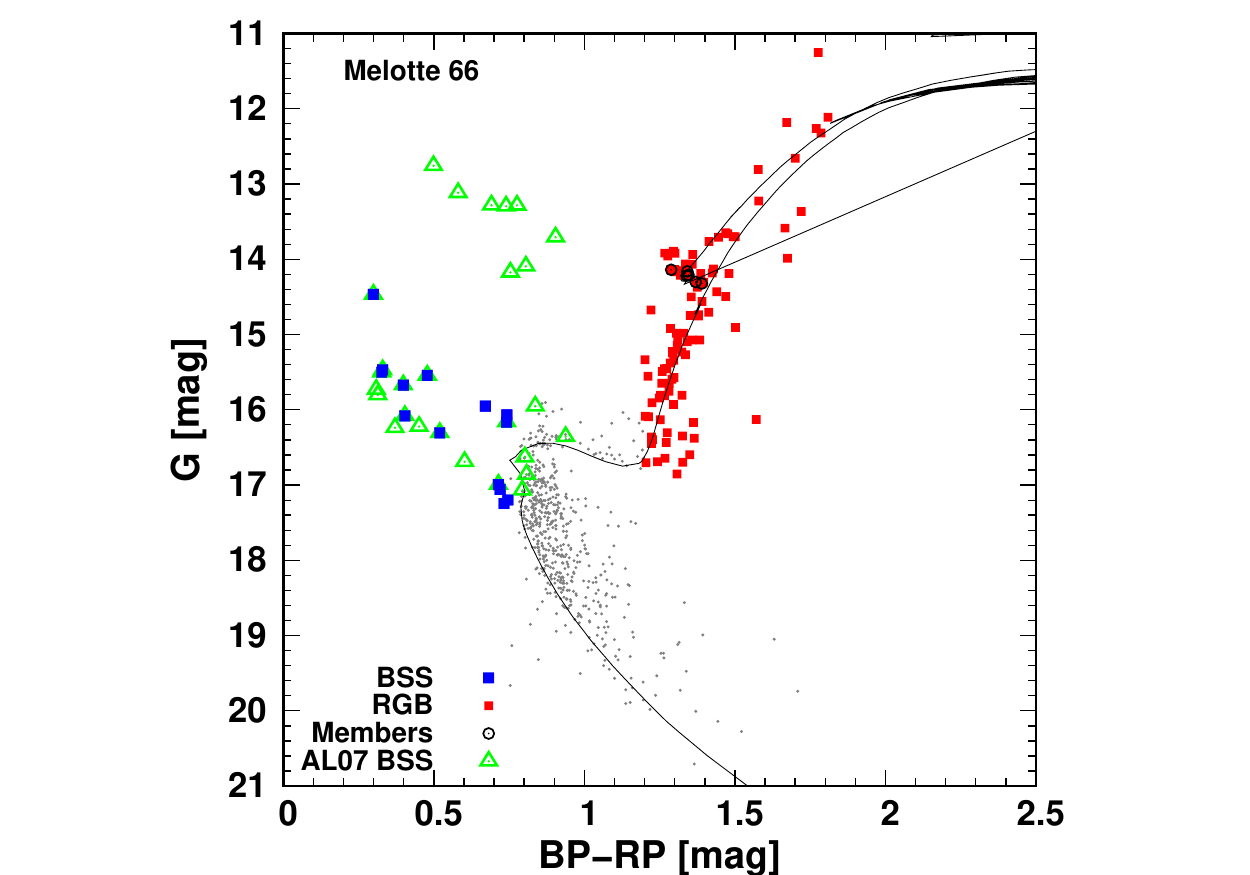}\par
\end{multicols}
\begin{multicols}{2}
\includegraphics[width=10cm]{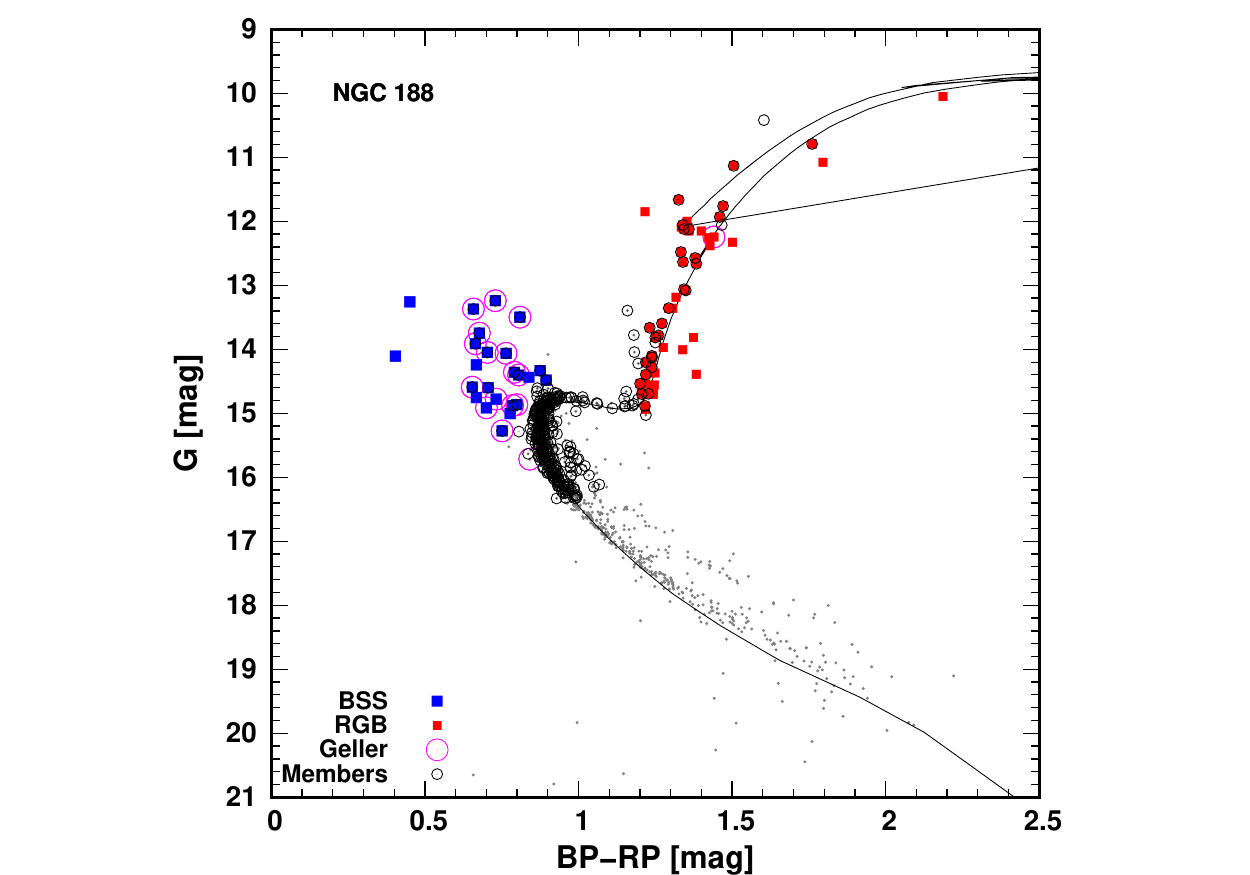}\par
\includegraphics[width=10cm]{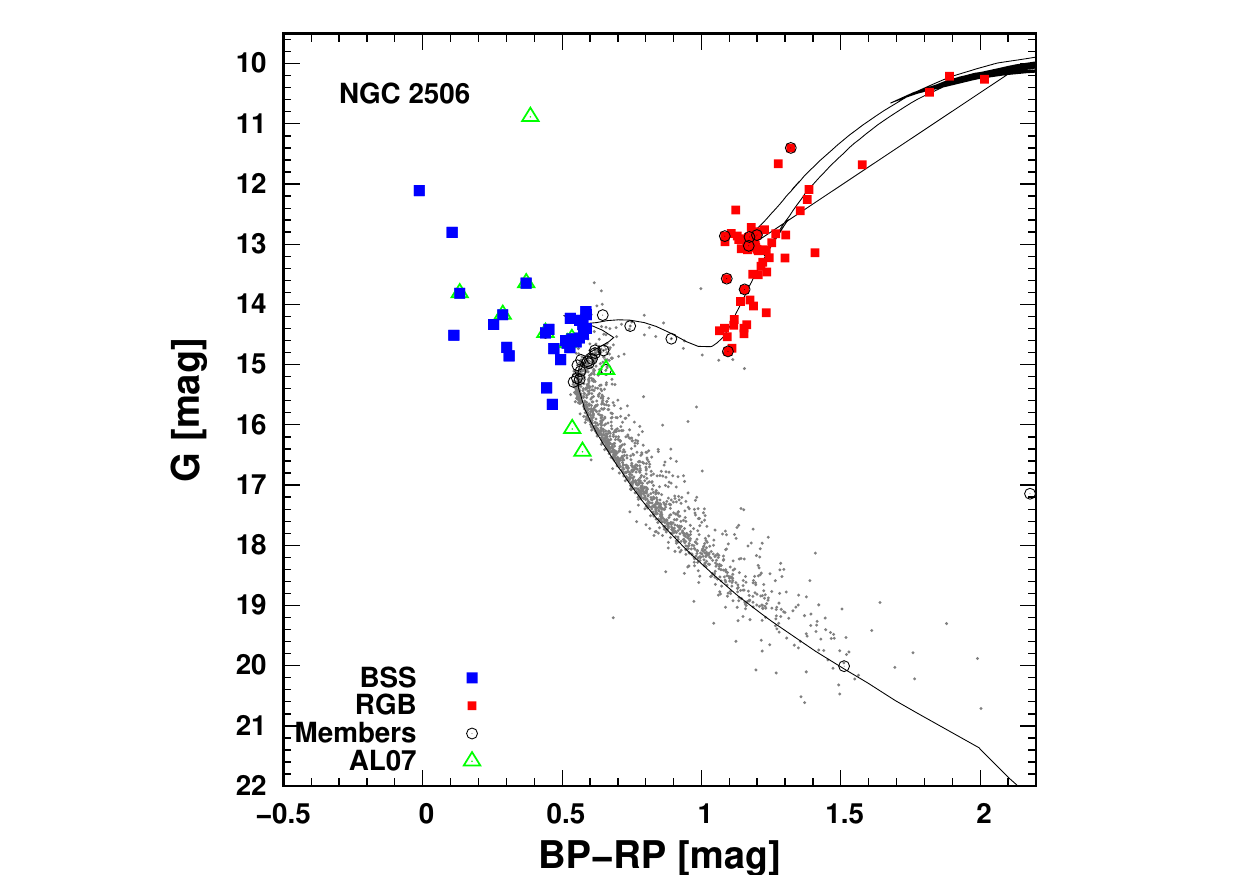}\par
\end{multicols}
\begin{multicols}{2}
\includegraphics[width=10cm]{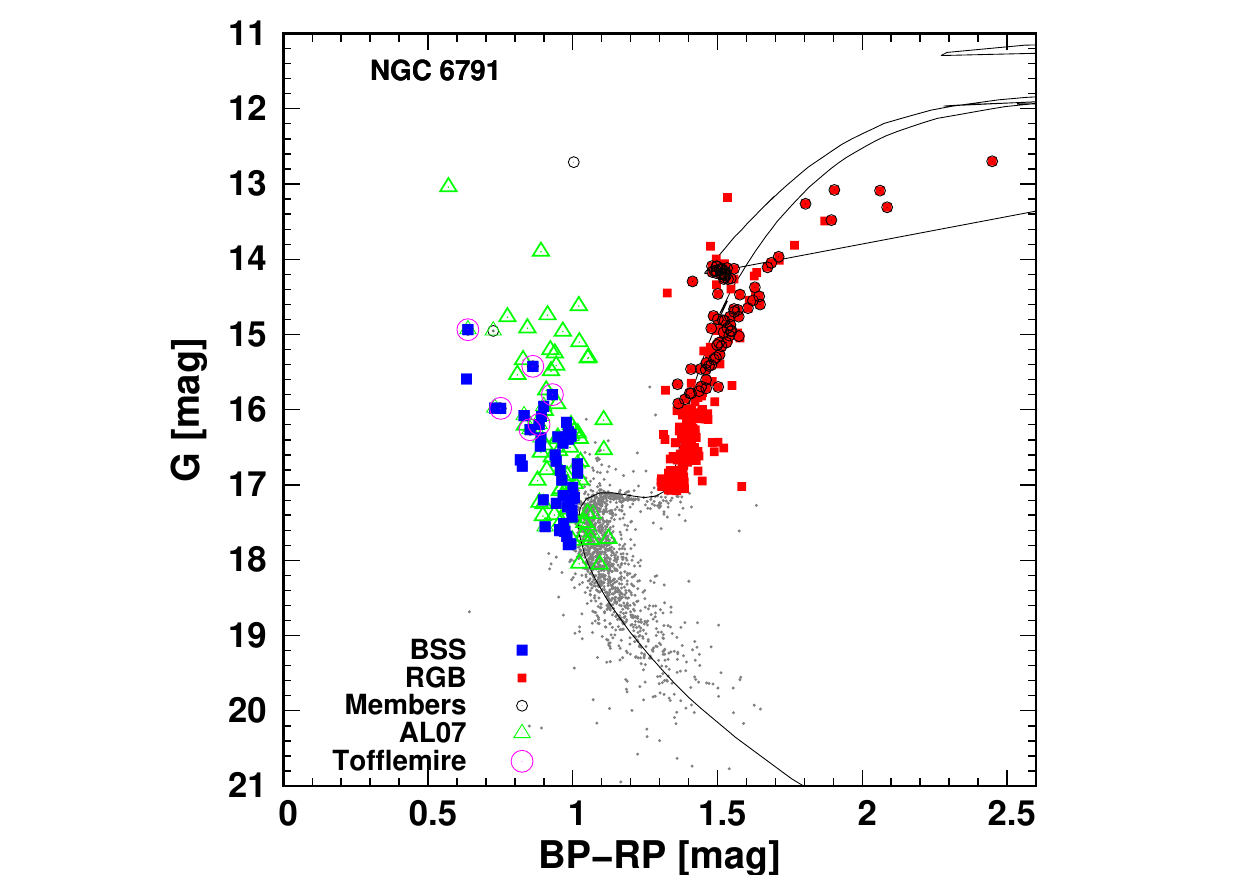}\par
\includegraphics[width=10cm]{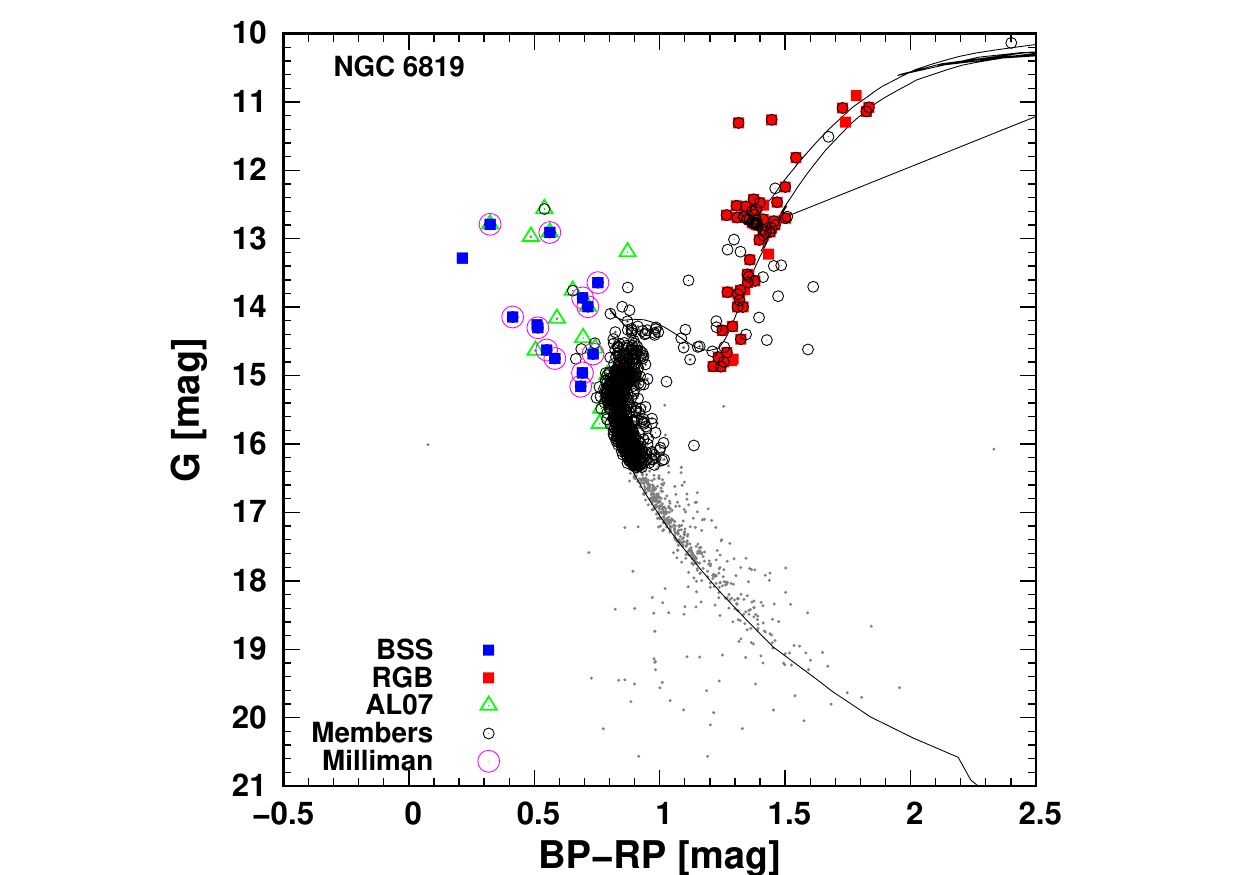}\par
\end{multicols}

 \caption{The CMDs of the clusters are shown with PARSEC isochrones. Our BSS marked as blue solid squares, BSS from AL07 are marked as green open triangles, and BSS known with spectroscopy data in the literature are marked as magenta open circles. Previously known spectroscopically confirmed members of the clusters are marked as black open circles.  Our RGB populations are marked as red solid squares.}
\label{Fig. A6}
 \end{figure*} 

\end{document}